\documentclass [preprint]{aastex}

\begin{document}

\title {\bf Resolving the Stellar Population of the Standard
Elliptical Galaxy NGC3379}

\author{Michael~D.~Gregg\altaffilmark{1,2},
Henry C. Ferguson\altaffilmark{3},
Dante Minniti\altaffilmark{4},\\
Nial Tanvir\altaffilmark{5}, and
Robin Catchpole\altaffilmark{6} 
}

\altaffiltext{1}{Physics Department, University of California, Davis,
CA 95616}

\altaffiltext{2}{Institute of Geophysics \& Planetary Physics,
Lawrence Livermore National Laboratory, Livermore, CA 94550}

\altaffiltext{3}{Space Telescope Science Institute, Baltimore, MD 21218}

\altaffiltext{4}{Depto de Astronomia, P. Universidad Catolica, Casilla
306, Santiago 22, Chile}

\altaffiltext{5}{Department of Physics, Astronomy, and Mathematics,
University of Hertfordshire, Herts, AL10 9AB, United Kingdom}

\altaffiltext{6}{Royal Observatory, Greenwich, London SE10 9NF}

\authoraddr{gregg@igpp.ucllnl.org, Institute of Geophysics \&
Planetary Physics, L-413, Lawrence Livermore National Laboratory,
Livermore, CA 94550}

\begin{abstract}

Using the Near Infrared Camera and Multi-Object Spectrometer (NICMOS)
on board the Hubble Space Telescope, we have obtained $F110W$ ($\sim
J$) and $F160W$ ($\sim H$) images of three fields in NGC3379, a nearby
normal giant elliptical galaxy.  These images resolve individual red
giant stars, yielding the first accurate color-magnitude diagrams for
a normal luminous elliptical.  The photometry reaches $\sim$ 1
magnitude below the red giant branch tip with errors of $\lesssim
0.2$~mags in $F110W-F160W$.  A strong break in the luminosity function
at $F160W = 23.68 \pm 0.06$ is identified as the tip of the red giant
branch (RGB); comparison with theoretical isochrones implies a
distance of $10.8 \pm 0.6$~Mpc, in good agreement with a number of
previous estimates using various techniques.  The mean metallicity is
close to solar, but there is an appreciable spread in abundance, from
at least as metal poor as [Fe/H]$\approx -1.5$ to as high as $+0.8$.
There is a significant population of stars brighter than the RGB tip
by up to $\sim 1$ magnitude.  The observations of each field were
split over two epochs, separated by $2-3$ months, allowing the
identification of candidate long period variables; at least $40\%$ of
the stars brighter than the RGB tip are variable.  Lacking period
determinations, the exact nature of these variables remains uncertain,
but the bright AGB stars and variables are similar to those found in
metal rich globular clusters and are not luminous enough to imply an
intermediate age population.  All of the evidence points to a stellar
population in NGC3379 which is very similar to the bulge of the Milky
Way, or an assortment of Galactic globular clusters covering a large
metallicity spread.

\end{abstract}

\keywords{elliptical galaxy: stellar populations}

\section {Introduction}

Knowledge of the present stellar content and star formation histories
of early type galaxies is essential for theories of galaxy evolution
and has important consequences for the use of ellipticals in the
distance ladder.  A major barrier to divining the nature of
luminous elliptical galaxies is the dearth of nearby examples whose
stellar populations can be resolved and studied star-by-star.
Consequently, considerable effort has been expended in spectral
synthesis modeling of the integrated light of early type galaxies
(e.g.\ Leitherer, et al.\ 1996), but spectral modeling has not yet
clarified the nature of the stellar population of normal giant
elliptical galaxies.  There is often significant disagreement over
even the most fundamental issues of mean age and metallicity among
various methods, as was documented by Arimoto (1996) in
a blind test of many spectral synthesis methods by independent
researchers.  Resolved stellar population analyses of elliptical
galaxies is essential for making real progress.

Ground-based color-magnitude diagrams (CMDs) in the near-infrared by
Freedman (1989; 1992) and Elston \& Silva (1992) for the Local Group
high surface brightness, compact dwarf elliptical M32 (NGC221)
appeared to exhibit an excess of stars well above the tip of the red
giant branch (RGB), evidence for an intermediate age population.  More
recent V-I CMDs from the Hubble Space Telescope (HST) using WFPC2
(Grillmair et al.\ 1996), however, find no direct evidence for this
population, apparently an artifact of crowding in the ground-based
studies, as was predicted by Renzini (1992).  After M32, the next
nearest early type galaxy stand-in is the spheroid of NGC5128.  Soria
et al.\ (1996), Harris et al.\ (1999), and Marleau et al.\ (2000) have
constructed CMDs using HST WFPC2 V + I or NICMOS J + H images of the
outer halo of NGC5128, detecting the tip of the RGB and plentiful
bright asymptotic giants, the latter strongly suggesting an
intermediate age population.  Maffei~1, at a distance of $\sim 4$~Mpc
(Luppino \& Tonry 1993),
is probably the nearest normal elliptical galaxy (Buta \& McCall
2003), but at a Galactic latitude of $-0\fdg55$, it is severely
obscured by large and highly variable Galactic extinction,
$A_V\approx5$, making it a very difficult object for study at optical
and IR wavelengths.  Using ground-based adaptive optics, Davidge
(2002) has resolved AGB stars in Maffei~1 in the $H$ and $K$ bands,
reaching $K\approx22$ with errors of $0.25$ magnitudes, but not yet
good enough to detect the RGB tip or study the RGB population in
detail.

These results are important for understanding early type galaxy
populations, but neither M32 nor NGC5128 is a normal, luminous,
elliptical galaxy.  With M$_{\rm V} = -16$, M32 is at the low
luminosity extreme of high surface brightness ellipticals and its
proximity to M31 has almost certainly influenced its development
(Faber 1973; Nieto \& Prugniel 1983; Bekki et al.\ 2001).  NGC5128 is
quite peculiar, a probable recent merger harboring an active galaxy
nucleus (Soria et al.\ 1996) and stars as young as 10~Myr (Rejkuba et
al.\ 2001; 2002).  Study of Maffei~1 is greatly complicated and
compromised by the variable foreground Galactic dust across its field.
Color-magnitude diagrams for these unusual objects cannot, without
further investigation, be considered representative of the class of
standard, luminous ellipticals.

The nearest normal giant elliptical galaxy which can be studied free
of complications is NGC3379 in the Leo~I
galaxy group.  It is an E0, with typical early type colors, Mg$_2$
index, and velocity dispersion (Davies et al. 1987).  It is well fit
by an R$^{\onequarter}$ law (de Vaucouleurs \& Capaccioli 1979) and
appears to have no large scale morphological peculiarities (Schweizer
and Seitzer 1992).  There is some evidence that NGC3379 could be a
face-on S0 (Statler \& Smecker-Hane 1999; Capaccioli et al.\ 1991),
but the evidence is equivocal.  It contains a tiny nuclear dust ring,
$R \approx 1\farcs5$, and small ionized gas ring, $R \approx 8$
(Pastoriza et al.\ 2000; Statler 2001).  This amount of visible
interstellar material is small compared to that in the 40\% of typical
bright ellipticals with gas and dust (Faber et al.\ 1997; Sadler \&
Gerhard 1985).  The Leo~I group, which also includes the S0 NGC3384
and the Sc NGC3389, is distinguished by its large partial ring of
neutral hydrogen (Schneider 1989).

The HST Extragalactic Distance Scale Key Project (Freedman et al.\
2001) arrives at a distance of $\sim 9.5$~Mpc ($m-M = 29.90\pm0.10$)
for the Leo~I group, somewhat lower than most other estimates (Tanvir
et al. 1999, Graham et al.\ 1997; Tonry et al. 1990; Sakai, Freedman,
\& Madore 1996; see Gregg 1997 for a short summary).  At this
distance, the resolved stellar population of NGC3379 is all but
impossible to study with present ground-based instrumentation.  Sakai
et al.\ (1996) used WFPC2 to obtain I-band images just deep enough to
locate the tip of the RGB in NGC~3379 in a field $5-6\arcmin$ from the
center.  While providing a distance estimate of $11.5\pm1.6$ Mpc ($m-M
= 30.30\pm0.14$), marginally consistent with the Key Project distance,
their single filter observations do not constrain the metallicity of
the RGB or probe the nature of the AGB.

Apart from the detection of the RGB tip (TRGB) by Sakai et al.,
everything that is known about the stellar population of NGC3379 has
been inferred from spectroscopy and photometry of its integrated
light.  Compiling the spectral line data from a number of recent
studies, Terlevich \& Forbes (2002) have derived age and metallicity
estimates for 150 nearby elliptical galaxies, using the stellar
population models of Worthey \& Ottaviani (1997).  The mean results
for NGC3379 indicate values of $t = 9.3$~Gyr and [Fe/H]~$=+0.16$,
[Mg/Fe]~$=+0.24$, consistent with its having a canonical old,
relatively metal-rich stellar population.  Rose (1985), however, has
presented arguments based on detailed analysis of integrated light
spectral line strengths, that luminous ellipticals, NGC3379 included,
have a ``substantial'' intermediate age component, $6-7$~Gyr old,
though the fraction is not quantified.  These conclusions from
integrated spectra are based on single-metallicity population models
and so must be viewed with some reservations; NGC3379, like most
ellipticals, exhibits line strength and color gradients (Davies,
Sadler, \& Peletier 1993), indicative of abundance variations
(Peletier et al.\ 1999) and hence a composite metallicity population.
A metal-poor component can plausibly account for the relatively young
turnoff age reported by Rose (1985).

Using HST with NICMOS Camera~2 (Thompson et al.\ 1998), we have
resolved the bright RGB and AGB stars in three fields in the halo of
NGC3379 in the $F110W$ ($\sim J$) and $F160W$ ($\sim H$) filters.
These images have yielded CMDs, the first such data for a normal,
luminous elliptical galaxy which probe below the RGB tip with some
precision.  Analysis of these data allow us to place constraints on
the mean metallicity of the giant stars and to compare the CMDs to
those for nearer, relatively well-observed early type populations,
such as the Milky Way bulge and globular clusters, as a first attempt
in understanding the star formation history of a typical bright,
normal elliptical.  By dividing our NICMOS observations into two
epochs separated by many weeks, we have identified candidate variable
stars in its halo, opening another window on elliptical galaxy stellar
populations.

The analysis is summarized as follows: the $F160W$ luminosity function
yields the TRGB apparent magnitude ($\S 5.1$).  Comparison of the CMD
to theoretical isochrones yields a mean abundance which must be near
solar $\S 6.1$), irrespective of age.  With the abundance determined,
the isochrones provide a distance estimate via the TRGB, again with
very little age sensitivity; the largest sources of error are the
choice of whose isochrones to use and the binning of the isochrone
points at the tip of the RGB, rather than the age or abundance of the
stars, or the photometric and statistical uncertainty in the TRGB
location.  With the distance estimate, a more detailed comparison of
the CMDs to the isochrones reveals a large abundance spread of $-2 <
[{\rm Fe/H}] < +0.8$ ($\S 6.1.3$) while no young or intermediate age
stars ($t \lesssim 5$~Gyr) are required.  The contribution of any
population with age $\lesssim 5$~Gyr is put on more quantitative
footing by the number of very bright AGB stars, which sets an upper
limit of $\sim 20\%$ coming from such a population ($\S 6.1.4$).
Comparison of the NGC3379 CMDs to the Milky Way Bulge and globular
clusters and NGC5128 supports this analysis ($\S6.2$).  While a large
age spread in the range 8 to 15~Gyrs cannot be ruled out by our data,
there is no compelling evidence for any significant contribution from
stars with ages $< 8$ Gyrs.

\section{Observations}

The locations of our three NICMOS Camera-2 fields are shown in
Figure~1, on the Digitized Sky Survey image of NGC3379.  The
positioning was driven by two factors.  First was to sample a large
range in distance from the center so as to be sensitive to any radial
variations in stellar population.  The fields are 3\arcmin\ (8.7~kpc),
4\farcm5\ (13.0~kpc), and 6\arcmin\ (17.5~kpc) from the nucleus.  The
half-light radius $R_e$ of NGC3379 is 54\arcsec, so these fields
are at 3.33, 5.0, and 6.67 $R_e$ with $B$-band surface brightnesses of
25.2, 26.4, and 27.2 mags/\sq\arcsec, respectively (Capaccioli et al.\
1990).  The somewhat scattered locations were chosen in part to
maximize the usefulness of the parallel STIS and WFPC2 observations.
Unfortunately, the parallel observations were not executed as
efficiently as we had hoped and are not deep enough to probe the
resolved stellar population of NGC3379.

\begin{figure}[t]
\plotone{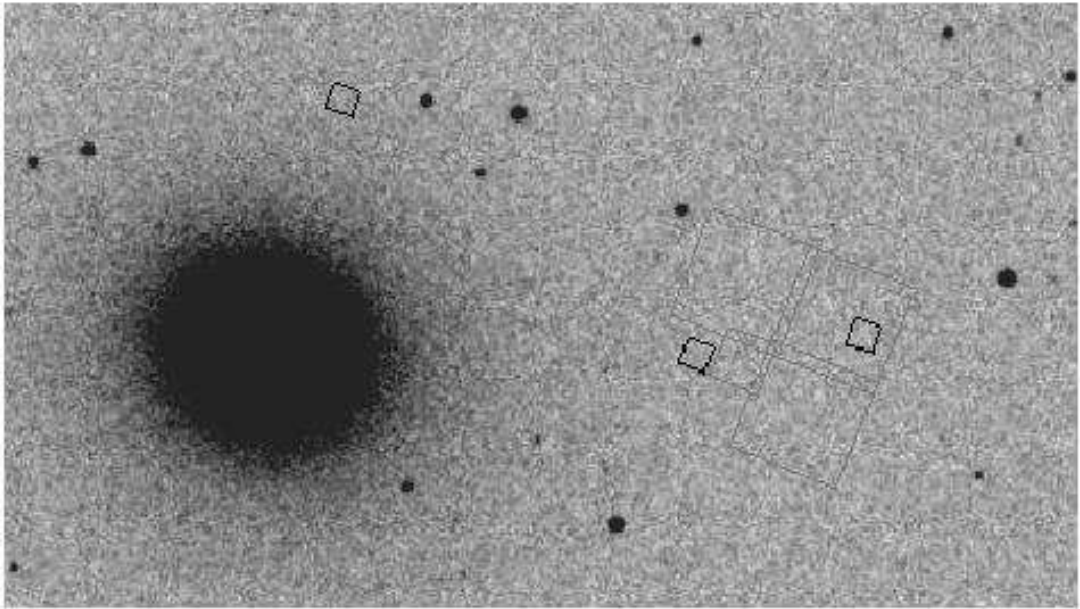}
\caption{\small Locations of the three NIC-2 fields (smaller dark squares) in
the halo of NGC3379.  At distances of 3\arcmin, 4.5\arcmin, and
6\arcmin from the nucleus, these fields have mean B surface
brightnesses of 25.2, 26.4, and 27.2, respectively.  The outermost field
is contained within the WFPC2 field of Sakai et al.\ (1996), also
indicated.}
\end{figure}

Observations of each of the three fields were divided evenly between
two epochs, spaced 2-3 months apart to identify candidate variable
stars.  The interval was set by HST scheduling constraints and the
desire to have the same orient for follow-up visits, while obtaining
all data in a single observing cycle.  A drawback in observing at two
epochs was the significant increase in zodiacal light during the
second visits.  The overall background was roughly twice as high,
completely consistent with the variation in zodiacal light due to the
object-HST-Sun viewing angle changing from $\sim 145\arcdeg$ to $\sim
56\arcdeg$.

Data were taken using $\sim 1400$s MULTIACCUM SPARS64 integration
sequences; the fields were dithered by a few pixels between sequences.
The total integration time spent in each filter for each field is
11.2~ksec (8 MULTIACCUM sequences).  We initially planned to divide
the exposure times between $F160W$ and $F110W$ with a 5:3 ratio.
After the first epoch observations, however, it was evident that we
were not obtaining sufficient depth in $F110W$, so the second visits
were used to equalize the total exposure times in the two filters.

\begin{deluxetable}{lllllllll}
\footnotesize
\tablewidth{0pt}
\tablecaption{NICMOS Observing Log 1998}
\tablehead{
\multicolumn{3}{c} {Field 1 (93 day interval)} &
\multicolumn{3}{c} {Field 2 (62 day interval)} & 
\multicolumn{3}{c} {Field 3 (63 day interval)} \\  
\colhead{Epoch}  & \colhead{Filter}  & \colhead{Int. time} &
\colhead{Epoch}  & \colhead{Filter}  & \colhead{Int. time} &
\colhead{Epoch}  & \colhead{Filter}  & \colhead{Int. time} 
}
\startdata
Apr 02 & $F160W$ & 1343.9 & May 02 & $F160W$ & 1343.9 & May 03 & $F160W$ & 1343.9 \\  
 &        &  1407.9  &   &        &  1407.9  &  &        &  1407.9 \\    
 &        &  1407.9  &   &        &  1407.9  &  &        &  1407.9 SAA \\
 &        &  1407.9  &   &        &  1407.9  &  &        &  1407.9  \\   
 &        &  1407.9  &   &        &  1407.9  &  &        &  1407.9 SAA \\
 &  $F110W$ &  1407.9  & &  $F110W$ & 1407.9 &  & $F110W$ &  1407.9  \\ 
 &        &  1407.9  &   &        &  1407.9  &  &        &  1407.9 \\    
 &        &  1407.9  &   &        &  1407.9  &  &        &  1407.9 SAA \\
Jul 04 & $F160W$ & 1407.9 & Jul 03 & $F160W$ & 1407.9 & Jul 05 & $F160W$ &  1407.9 \\  
 &        &  1407.9  &   &        &  1407.9 saa & &        &  1407.9 SAA \\
 &        &  1407.9  &   &        &  1407.9 SAA & &        &  1407.9 saa \\
 &  $F110W$ &  1343.9  & &  $F110W$ & 1343.9 &    & $F110W$ &  1343.9 \\  
 &        &  1407.9  &   &        &  1407.9  &    &        &  1407.9 \\    
 &        &  1407.9  &   &        &  1407.9  &    &        &  1407.9 \\    
 &        &  1407.9  &   &        &  1407.9  &    &        &  1407.9 saa \\
 &        &  1407.9  &   &        &  1407.9 SAA & &        &  1407.9 SAA \\
\enddata
\tablecomments{``SAA'' means the data were judged too badly affected
by SAA CR persistence to be useful; ``saa'' means
that the data show signs of CR persistence, but
are still included in the final images.}

\end{deluxetable}

Only the innermost field is uncompromised by cosmic ray (CR)
persistence during South Atlantic Anomaly (SAA) passage.  Severely
affected data were omitted from the final images, resulting in reduced
total exposure times.  The combination of higher background and CR
persistence has rendered the second epoch $F160W$ images for Field 2
nearly useless.  The outer field which overlaps with the previous
WFPC2 data is also seriously impacted, with effective exposures of
only 6975s in $F160W$ and 8383s in $F110W$, and even these have
noticeable CR persistence.  A log of the observations 
is given in Table~1.

\section{Data Reduction}

\begin{figure}[t]
\epsscale{0.75}
\plotone{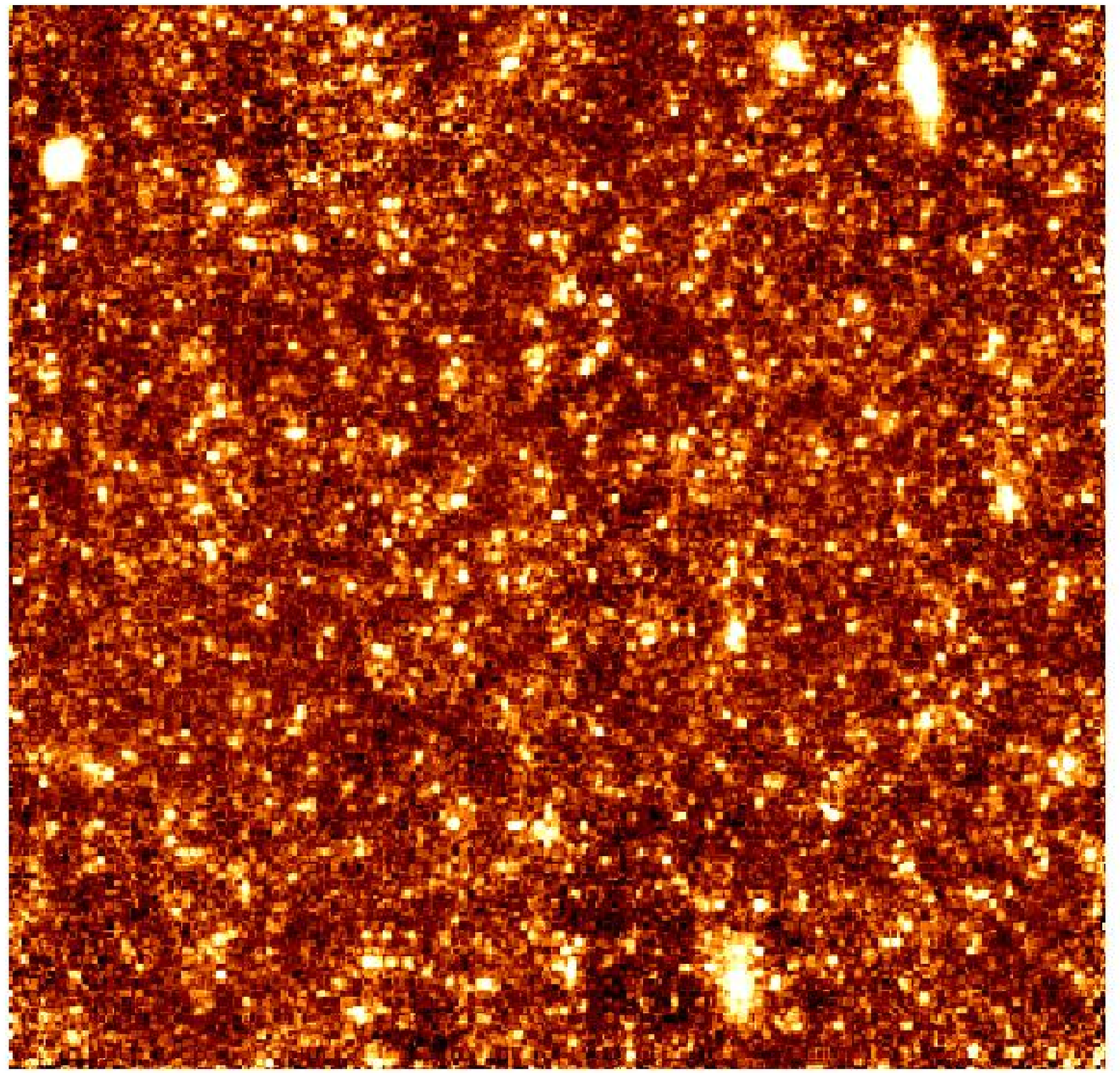}
\caption{\small Final drizzled, combined $F160W$ (H band) image of the
innermost 
field in the halo of NGC3379, 3\arcmin\ from the nucleus.
Total integration time is 11.2ks.
These data were little affected by SAA passages.  One bright Galactic
foreground star is present in the upper left corner.}
\end{figure}

\begin{figure}[t]
\epsscale{0.75}
\plotone{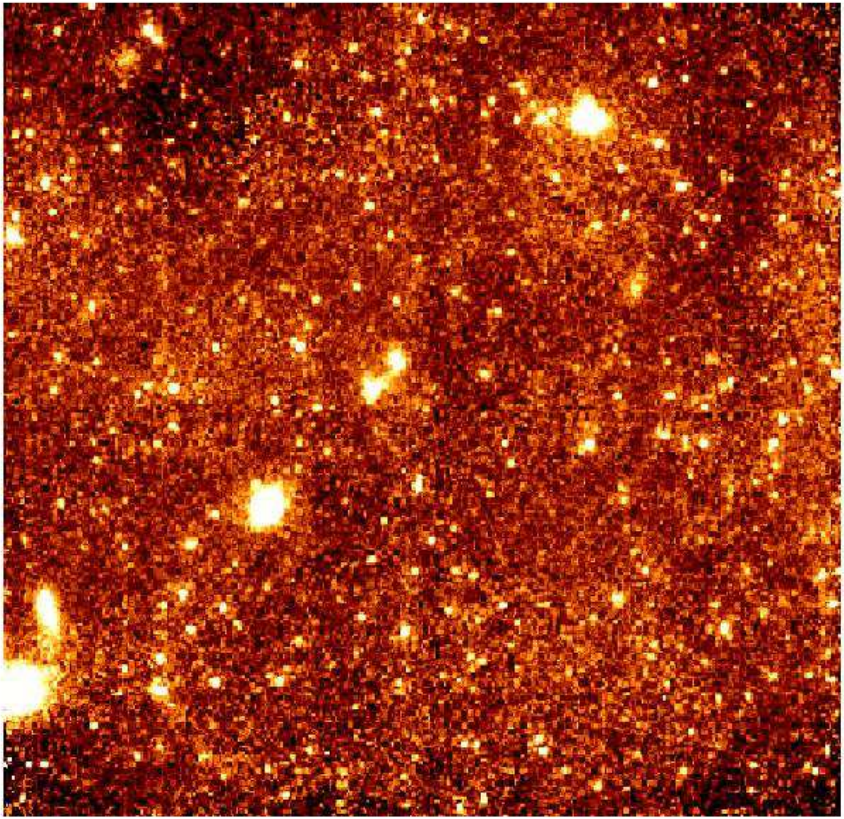}
\caption{\small Final drizzled, combined $F160W$ (H band) image of the middle
field in the halo of NGC3379, 4.5\arcmin from the nucleus.
The total exposure time is 9.8ks; one {\sc MULTIACCUM} sequence being
badly affected by SAA passage.  A number of extended background
sources contaminate the field.}
\end{figure}

\begin{figure}[t]
\epsscale{0.75}
\plotone{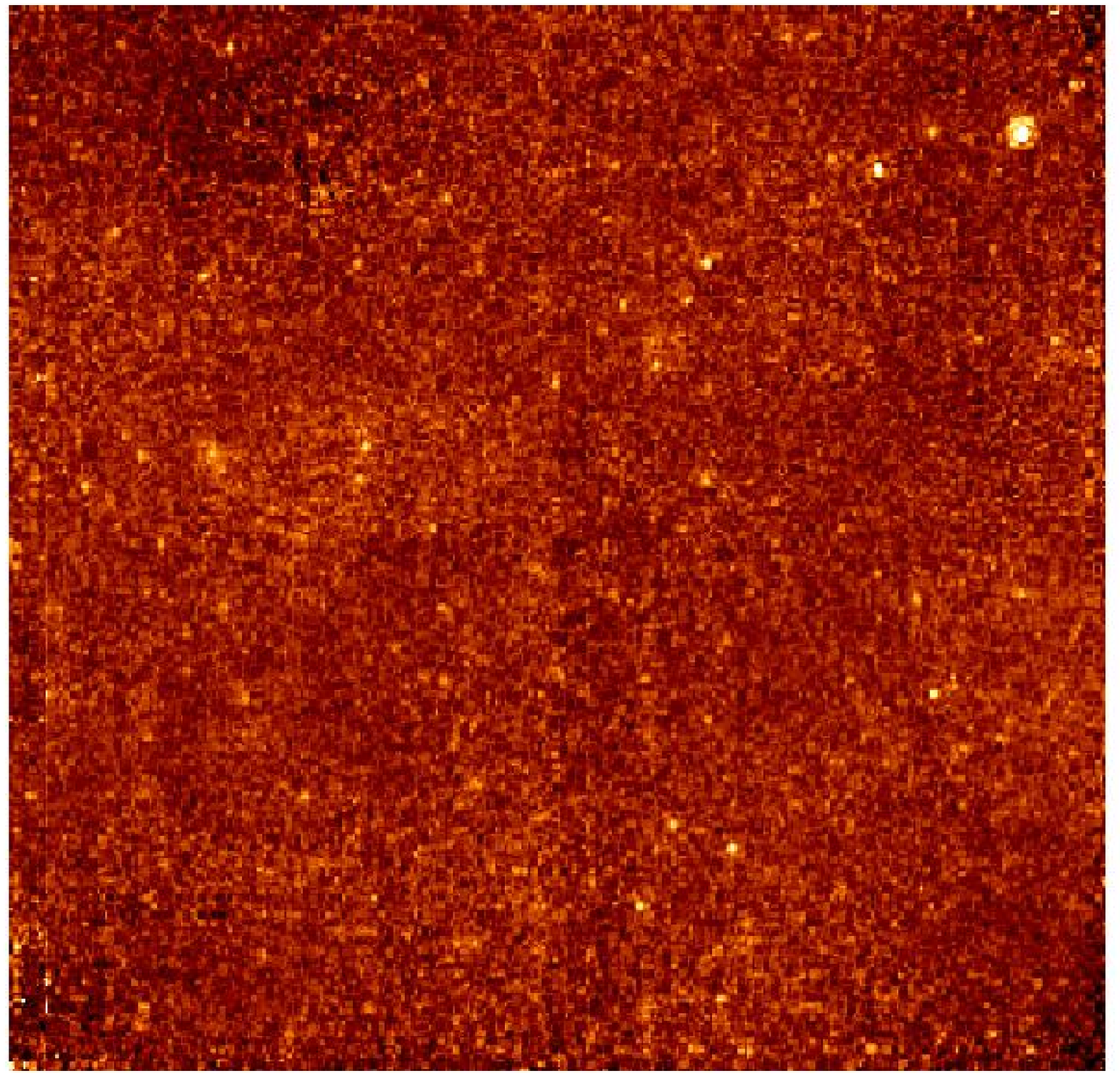}
\caption{\small Final drizzled, combined $F160W$ (H band) image of the
outermost field in the halo of NGC3379, 6\arcmin from the nucleus.
The total exposure time is 7.0ks; several {\sc MULTIACCUM} sequences
being badly affected by SAA passage.  One bright Galactic foreground
star is present in the upper right corner.}
\end{figure}

The raw NICMOS images were reduced using IRAF scripts kindly provided
by M.~Dickinson (1998, private comm.)  These scripts are very similar
to the now-public versions released in mid-1999 which improve on the
original CALNICA pipeline procedures, primarily in the removal of the
quadrant-dependent bias level or ``pedestal'' signature.  Updated dark
frames which included the modeled temperature dependent variable bias
level (``shading'') correction were supplied by E. Bergeron.  A
correction was also made to remove excess shot noise from non-optimum
dark subtraction; this was done by subtracting a noise-free map of the
difference between the non-optimal pipeline dark correction and a more
accurate, very deep blank field frame; this ``delta-dark'' was also
supplied by M.~Dickinson.

The 8 reprocessed individual {\sc MULTIACCUM} exposures in each filter
were then registered using {\sc xregister} in {\sc IRAF}.  At this
point it was possible to assess the impact of the SAA on individual
exposures.  The unaffected {\sc MULTIACCUM}s were then combined using
the ``drizzle'' technique (Fruchter \& Hook 2002) to produce images
subsampled by a factor of 2 using a 0.9 drop size.  A very low order smooth
surface was fit to the individual frames and subtracted to remove the
background to bring all the frames to the same
mean value, necessary to achieve good results when drizzling.  A mean
sky level was added back to the final frame to preserve background
counts for computing photometric errors.  To look for variable stars,
separate images were made for each of the two epochs in the same
manner.

The final drizzled $F160W$ images are shown in Figure~2-4.  The
substantial gradient in stellar density from field to field is
immediately apparent.  A string of background galaxies is visible in
the middle field.  The outermost field is of very limited value
because it is so sparsely populated with stars, partly because of its
extreme outer location but also because it is most impacted by the
higher noise levels and reduced effective exposure times due to SAA CR
persistence.  This is quite unfortunate because this field overlaps
with the previous WFPC2 $I$-band observations.  This NICMOS location
detects so few stars, in fact, that comparison with the optical data
is useless.

\section{Photometry}

\subsection{PSF Fitting and Calibration}

Point-spread-function (PSF) fitting photometry was performed using
standard {\sc iraf/daophot} procedures, after masking out obvious
background galaxies and the NICMOS coronograph hole.  Attempts to
construct a reliable PSF model from the data were not successful
primarily because crowding raised the noise level in the wings of the
bright, well-exposed stars to unworkable levels.  Instead, we used the
{\sc tiny tim} package (Krist 1993) to construct NICMOS $F110W$ and
$F160W$ PSFs.  We added ``jitter'' to the {\sc tiny tim} models until we
reproduced the well-determined cores of the empirical PSFs in the
final drizzled images.  Because of the better-behaved profile wings,
the model PSFs gave much better results when used in the {\sc daophot}
routines, judging from the lower residuals in the subtracted images
and the much lower photometric errors.

We used apertures of 2.5 ($F110W$) and 3.0 ($F160W$) pixels radius to
measure the instrumental magnitudes in each drizzled image.  Objects
were found independently in $F110W$ and $F160W$ and then matched with
a cutoff distance of two drizzled pixels, allowing for possible small
registration errors, geometric distortions, and noise.  Using modeled
PSFs for the photometry made determination of the aperture corrections
straightforward; from the measurement aperture to a 0\farcs5 radius
(13.33 drizzled pixels) required an $F110W$ correction of $-0.65$
magnitudes while the $F160W$ correction is $-0.78$.  These magnitudes
were then corrected to ``infinite'' apertures with the standard factor
of 1.15.  We adopted the NICMOS Data Handbook ``Vega magnitude''
zeropoints of 1775.0 mJy in $F110W$ and 1040.7 mJy in $F160W$.  The
photometry has also been corrected for the small Galactic extinction
towards NGC3379, 0.022 and 0.014 magnitudes in $J$ and $H$ (Schlegel,
Finkbeiner, \& Davis 1998).

To ensure that only the best measured stars are used in the analysis,
the photometry was then filtered against several criteria reported by
the PSF fitting task.  Objects with $\chi^2 > 2.5$, ``sharpness''
parameter outside the range $-0.9$ to $1.$, and photometric error $ >
0.5$ in $F110W-F160W$ were discarded.  The first two criteria
eliminate nearly all extended objects, residual bad pixels, and edge
effects.  Total numbers of objects with both $F110W$ and $F160W$
photometry meeting the above criteria are 1751, 477, and 144 in Fields
1, 2, and 3, respectively.  The statistics of the photometric errors
are reported in Table~2.

\pagebreak
\begin{deluxetable}{cccccc}
\tablewidth{0pt}
\tablecaption{Photometry Statistics}
\tablehead{
\multicolumn{2}{c}{$F160W$}&\colhead{}  & \multicolumn{3}{c}{$F110W-F160W$}\\
\cline{1-2} \cline{4-6}\\
\colhead{Mag}  & 
\colhead{$<Error>$}  & 
\colhead{} &
\colhead{Mean}  & 
\colhead{$<Error>$}  & 
\colhead{RMS} }
\startdata
   22.50 &  0.020 & &   1.350 &    0.032  &     0.22 \\
   23.00 &  0.024 & &   1.436 &    0.048  &     0.21 \\
   23.50 &  0.033 & &   1.362 &    0.056  &     0.21 \\
   24.00 &  0.048 & &   1.358 &    0.080  &     0.24 \\
   24.50 &  0.072 & &   1.275 &    0.111  &     0.30 \\
   25.00 &  0.107 & &   1.174 &    0.161  &     0.43 \\
\enddata
\tablecomments{The RGB tip is at $F160W=23.68$.}
\end{deluxetable}

\subsection{Identification of Variable Star Candidates}

\begin{figure}[t!]
\epsscale{0.8}
\plotone{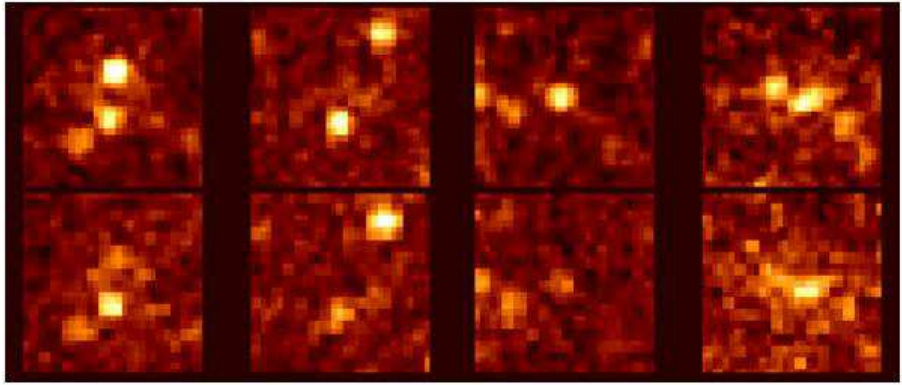}
\caption{\small Four examples of variable star candidates from the inner $F160W$
field are displayed with the brighter epoch above.  The first pair is
the most significant variable in the images with $\Delta F160W > 1.1$,
fading from $22.9$ to $24.0$ during the 3 month interval between
observations.   Both of the bright
stars in the second pair of images are identified as variable.  The
rightmost is a fainter example, with $<F160W> = 24.1$ and an amplitude of 0.6
magnitudes. 
}

\end{figure}
\begin{figure}[ht!]
\vspace{-0.25in}
\epsscale{0.6}
\plotone{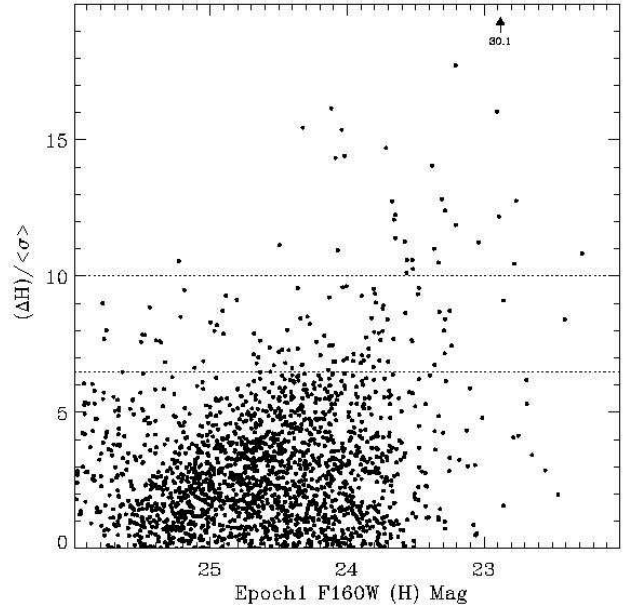}
\caption{\small Variable star identification was done by comparing
the change in magnitude between epochs relative to the photometric
errors for every object identified in both epochs of $F160W$
images.  Because stars may vary below the detection threshold in one
epoch, this approach may miss some of the most extreme
variables, but all stars with $F160W < 24$ were recovered in both epochs.
The variable with the highest significance is off the top of the plot
(arrow).  }
\end{figure}

\begin{deluxetable}{cccc}
\tablewidth{0pt}
\tablecaption{Field 1 Variable Star Statistics}
\tablehead{
\colhead{F160W range}  & 
\colhead{N$_{\rm TOT}$}  & 
\colhead{N$_{\rm VAR}$}  & 
\colhead{F$_{\rm VAR}$}  }
\startdata
$22.2-23.65$ &  \phn \phn 98 & 38 &   0.388   \\
$23.65-24.0$ & \phn   197 &    24 &   0.122   \\
$24.0-24.5$  & \phn   503 &    33 &   0.066   \\
$24.5-25.0$  & \phn   569 &    29 &   0.051   \\
$25.0-25.5$  & \phn   329 & \phn 2 &  0.006   \\
\hline
$22.2-25.5$  &   1696 &   121 &   0.071
\enddata
\tablecomments{The RGB tip is at $F160W=23.68$}
\end{deluxetable}

The observations were split between two epochs for the express purpose
of detecting variable stars.  The intervals for fields 1, 2, and 3,
are 93, 62, and 63 days, respectively.  Blinking the reduced images
immediately revealed many variables (Figure~5).  To quantitatively
identify and characterize the variable star candidates, we separately
ran {\sc daophot} on the reduced and combined images for each epoch.
The positions of stars determined in the total integration time images
were used to constrain the photometry for the noisier separate epoch
images.  With just two epochs, the problem of identifying variable
star candidates reduces to gauging the significance of the change in
brightness of a star from first to second epoch, which can be done in
myriad ways.  For Field 1, we show $\Delta(H) = (F160W_1 - F160W_2)$
divided by the photometric errors summed in quadrature from each
epoch, plotted against the first epoch $H$ magnitude (Figure~6).
Because stars may vary below the detection threshold in one epoch,
some of the most extreme variables are potentially missed, but, as it
happens, all of the stars in the combined image brighter than $F160W
\approx 24$ were detected in both epochs.

The dotted lines at the 6.5 and 10 ``sigma'' levels are drawn to help
evaluate the diagram, but can also be taken as liberal and
conservative criteria for variability.  In Field~2, no variables were
reliably detected because the second epoch images are so
shallow due to SAA problems.  A few stars qualified as variable in
Field~3. 

Many of the brightest stars with the tiniest photometric errors are
among the most significantly variable (Figures~5, 6), providing
confidence that the majority of the brightest stars are not due to
crowding of ordinary red giants.  The fractions of variables as a
function of $F160W$ magnitude are listed in Table~3.

\section{Luminosity Functions}

The $F160W$ luminosity function (LF) histograms for the combined epoch
imaging data for Fields~1 and 2 are shown in Figure~7; bin size is 0.1
magnitudes.  Field~3 has such poor statistics -- no stars at all
within 0.25 magnitudes of what will turn out to be the RGB tip -- that
we will not discuss its LF further.  The Field~1 and 2
distributions both begin to turn over in the $F160W = 24.7$ bin, suggesting
that this is where incompleteness becomes important.  This is
confirmed by the artificial star tests which indicate $\sim 80\%$
completeness at this brightness (see Appendix~A).  The LF for Field~1
(shown with and without the variable candidates) shows a definite jump
in the number of stars in the bin spanning $23.65 - 23.75$, which we
attribute to the brightness limit of the helium-burning RGB tip.
Omitting the variables, nearly all of which are probably AGB rather
than RGB stars (see $\S$6.1.4), increases the significance of this break
in the luminosity function; the RGB tip bin has more than 2.5 times as
many non-variables as the next brighter bin.

\begin{figure}[h!]
\vspace {-0.25in}
\epsscale{0.95}
\plotone{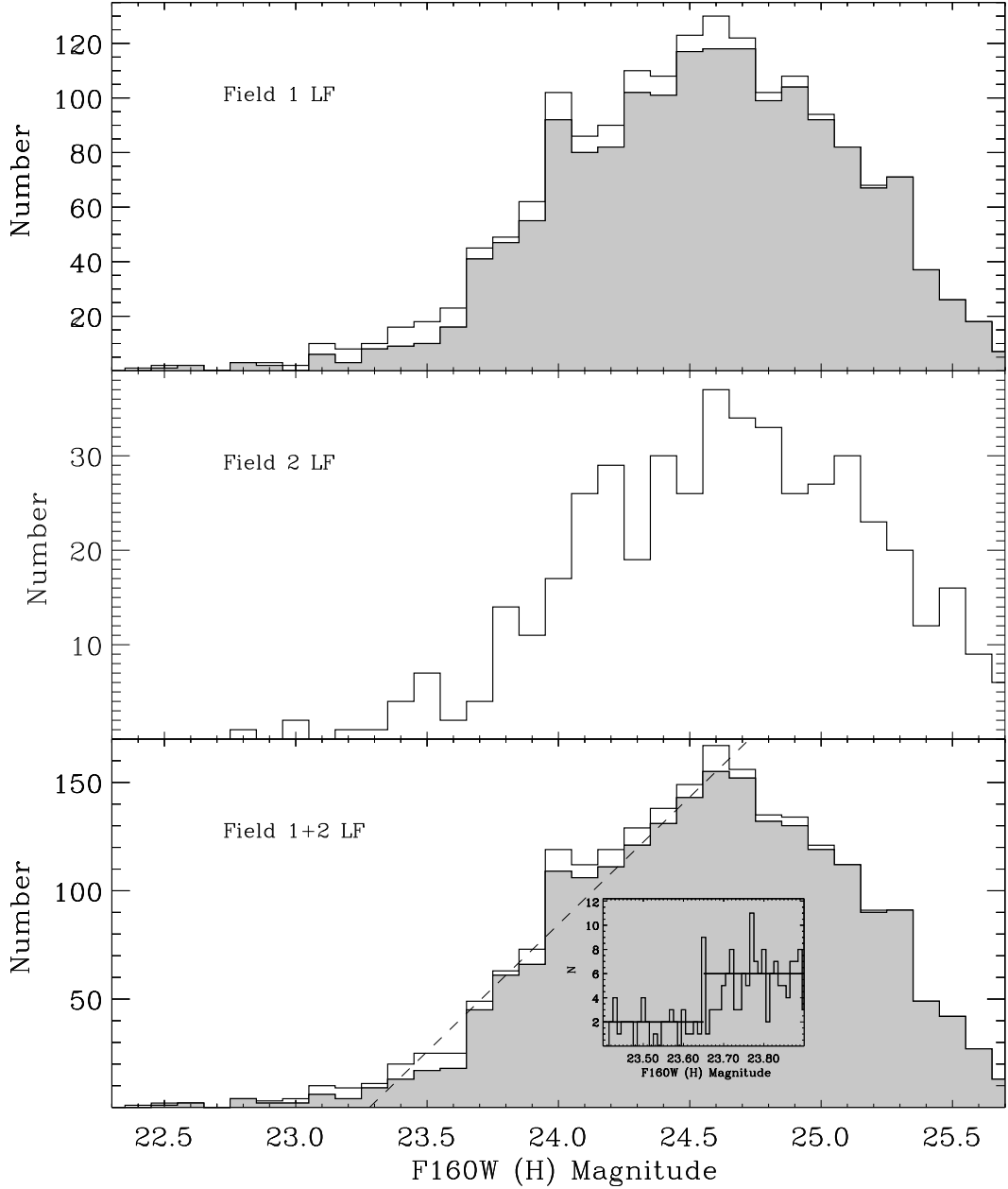}
\caption{\small The luminosity functions of the two inner NICMOS fields, and
their sum.  The unshaded portions of the upper and lower panel
histograms show the contribution from variable star candidates.  The
clear break at $F160W = 23.65$ is due to the tip of the RGB; a
higher resolution close-up of this region is displayed in the lower
panel.  Maximum-likelihood fitting yields a best-fit TRGB magnitude of
23.68.  The feature at $F160W = 24.0$, either an excess or a change in
slope, may be due to an AGB star contribution.  The dashed line is a
robust linear fit to the nonvariable star bins with $23.65 \leq F160W
\leq 24.6$  }
\end{figure}

The Field~2 distribution shows a similar but not as prominent break in
the next fainter bin, $23.75 - 23.85$ magnitudes.  As both fields are
at the same distance, this shift, if real, could indicate a modest
metallicity difference of $\sim 0.1$~dex (see $\S 6$), but with the
small number of stars in Field~2, this is not highly significant.  The
summed LF is shown in the bottom panel; the two histograms again being
with and without the variables from Field~1.  In the combined LF, the
significance of the RGB tip break remains roughly the same as for
Field~1 alone (2.50 times as many nonvariable stars as the next
brighter bin), even though we are unable to flag variables in Field~2.
The slope of the LF is close to linear over the brightest magnitude of
the RGB, as shown by the dashed line fit to the nonvariables in
Figure~7.  This fit predicts that the first bin brighter than the RGB
tip should contain more than twice as many stars as observed (37 vs.\
18).

\subsection{Pinpointing the Magnitude of the Tip of the RGB}

A blow-up histogram of the break region is shown in the inset; the bin
size is reduced to 0.01 magnitudes.  Even at this fine scale, there is
an obvious increase in the numbers of stars in each bin beginning at
$F160W=23.65$.  In the inset, the median number of stars per bin
brighter than this is 2; fainter is 6, firmly placing the RGB tip
magnitude at a level where the formal photometric errors are only $\pm
0.04$ magnitudes per star.  There are about a dozen stars within 0.01
magnitudes of the tip of the RGB, so the error budget for the RGB tip
magnitude is dominated by the systematic uncertainty in the
photometric calibration of NICMOS -- $5\%$ -- over any other source of
error.  Artificial star tests show that crowding effects for Field~1
introduce a median systematic error of only $\sim 0.01$ magnitudes per
star at this brightness (Appendix A).  In the face of 0.04 magnitude
errors, the extremely sharp break at $F160W=23.65 \pm 0.01$ must be
partly fortuitous, but certainly suggests an unambiguous placement of
the RGB tip.
To obtain a completely objective estimate of the RGB tip apparent
magnitude, we employed a maximum likelihood approach.  Figure~8 shows
the composite field 1+2 luminosity function (excluding the identified
variable stars) on a logarithmic scale with 0.04 magnitude binning.
We have determined the magnitude of the TRGB by fitting a two
power-law model:

\[ N(f) = \left\{ \begin{array}{ll}
           f^{\gamma_1} & \mbox{for $f > f({\rm TRGB})$};\\
           A f^{\gamma_2} & \mbox{for $f < f({\rm
           TRGB})$}. \end{array} \right. \]

The free parameters are $f({\rm TRGB})$, the flux corresponding to the
TRGB magnitude, the amplitude of the TRGB discontinuity, $A$, and the
two power-law indices $\gamma_1$ and $\gamma_2$.  To constrain the
models, we perform a maximum-likelihood fit on the {\it unbinned}
magnitudes.  As a simplification in our maximum-likelihood fit, we have
not convolved the model with the magnitude errors.  At the TRGB the
median error is 0.035 mag.  We verify through Monte-Carlo simulations
that this simplification does not significantly bias the results (see
below).  For each set of model parameters we compute the probability
distribution of star fluxes.  We then vary the parameters of the model,
maximizing the log of the likelihood of observing the data.  Fitting
data in the range $22.5 < F160W < 24.5$ (905 stars) we find a best-fit
TRGB magnitude of F160W = 23.68.  The best-fit model is shown as a
solid line in Figure~8.

\begin{figure}[t!]
\vspace {-1.0in}
\epsscale{0.9}
\plotone{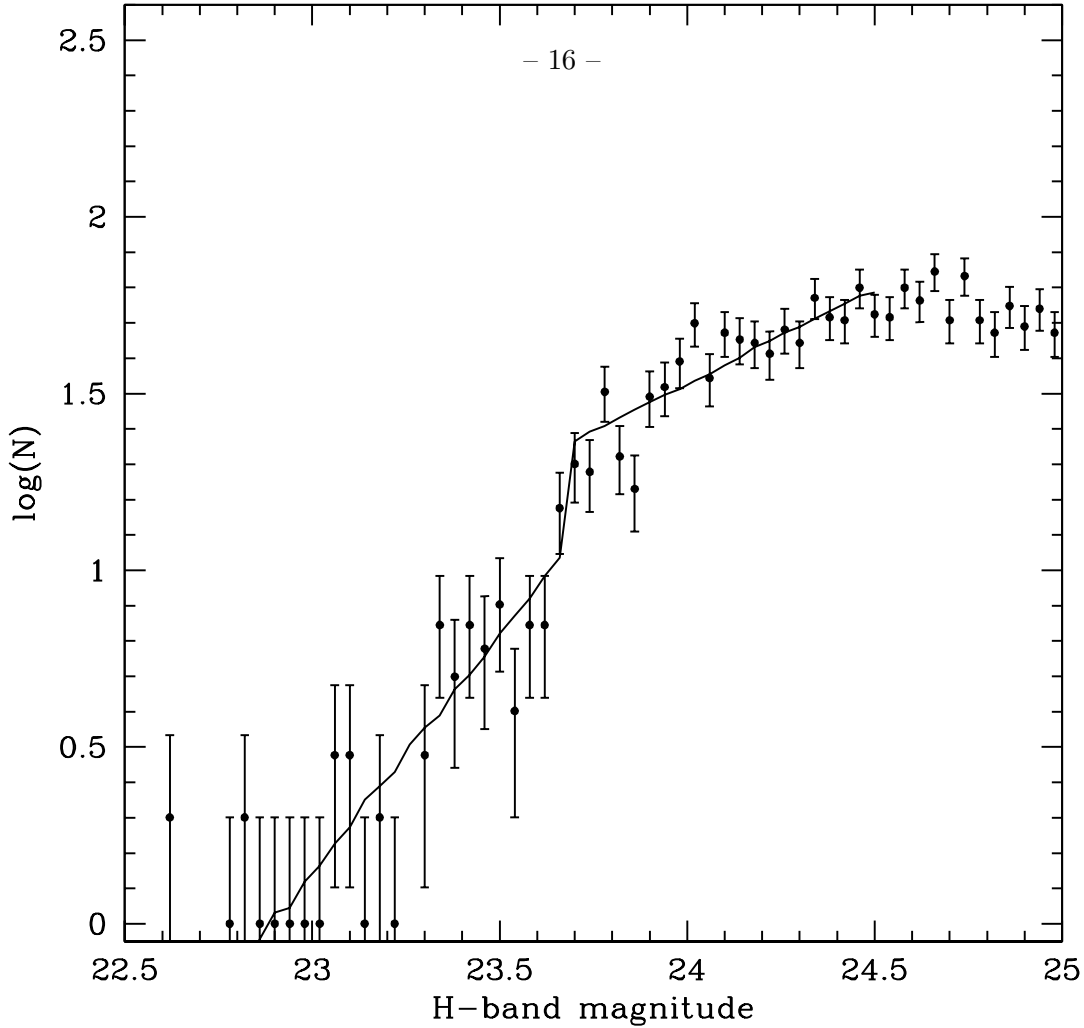}
\caption{\small The composite LF of Fields 1 and 2 (excluding the identified
variable stars) on a logarithmic scale with 0.04 magnitude binning.
The {\em unbinned} data for the 905 stars in the
range $22.5 < F160W < 24.5$ are modeled with two power laws using a
maximum-likelihood fit.  The best-fit model (solid line) yields a
TRGB magnitude $F160W = 23.68$ with 68\% confidence interval of $
23.63 < F160W < 23.70$.
}
\end{figure}

We have assessed uncertainties in two ways: (1) by bootstrap
resampling the data and re-performing the fit and (2) by creating
simulated data sets with realistic magnitude errors and known LFs and
carrying maximum-likelihood fits to verify that the input parameters
can be recovered.
The 68\% confidence interval spans the range $ 23.63 < F160W < 23.70$
and the 95\% confidence interval is $23.62 < F160W < 23.76$.  The
parameters $\gamma_1$ and $\gamma_2$ have been allowed to vary in
these experiments --- the best-fit TRGB magnitude is remarkably
robust.

The Monte-Carlo simulations allow us to determine if the fitting
procedure introduces any bias.  In our simulations, data points are
drawn from a model distribution function and scattered with a
magnitude-dependent error derived from the real data data: $\sigma(m)
= 10^{0.03449 m - 9.605}.$  Ten thousand data sets with the same number
of data points as the observed sample are created and best-fit model
parameters determined using the identical maximum-likelihood
procedure that was applied to the real data.  We tried input models
with a range of parameters similar to the true data.  As a typical
example, for a model with parameters $m({\rm TRGB}) = 23.65, A = 2.0,
\gamma_1 = -2.3$ and $\gamma_2 = -4.3$, we recover in 10000
Monte-Carlo iterations $m({\rm TRGB}) = 23.64 \pm 0.02.$  Ignoring the
magnitude errors in performing the maximum-likelihood fits thus does
not appear to introduce any significant bias in the results.  The
dominant source of error in the RGB tip magnitude is likely to be the
5\% systematic uncertainty in the NICMOS photometric calibration.
Departure of the true shape of the LF relative our the simple model
of a sharp discontinuity --- due to binaries, stellar rotation,
variability, and the finite metallicity spread, for example --- is
another possible source of systematic error for the TRGB-distance
technique.

While in principle the LF in $F110W$ might provide additional
information on the TRGB location, the tip is much less pronounced in
$F110W$.  In $\S6$, it will be shown that NGC3379 has a large
abundance spread.  Theoretical isochrones show that the TRGB is a
strong function of abundance in the $F160W$ band, with the
solar and near-solar metallicity TRGBs rising well above the more
metal-poor TRGBs, effectively producing a nearly single-abundance TRGB
to measure.  In $F110W$, however, the RGB tips of different
metallicity populations are closer together in luminosity, and the
large abundance spread of NGC3379 blurs the discontinuity so evident
at $F160W = 23.68$ (Figure~7).

\subsection{Other Luminosity Function Details}

In Field~1, there are 98 stars in the bins brighter than the tip; 38
of these (39\%) are variable candidates.  These will be discussed
further in the next section, along with the implications for the
distance and stellar population of NGC3379.  Also in Field~1, there is
an excess of stars in the bin centered at $F160W = 24$, which, in the
summed LF becomes another apparent break: the LF jumps by a factor of
1.65, then falls again.  This excess is not as statistically
significant as the brighter RGB tip break, and occurs in just one or
two bins.  Taking the average number of nonvariables in the bins on
either side of this feature predicts 85 stars whereas 109 are seen,
about a $2.5\sigma$ difference.  This second break can be interpreted
as an AGB contribution and is discussed further in $\S 6.1.4$.

\section{Color Magnitude Diagrams}

The NICMOS CMDs for each of the three fields are compared in Figure~9.
The variable star candidates identified in Figure~6 from Fields 1 and
3 are circled, double circles indicating the variables with
significance levels $> 10$.  The colored solid lines
trace the smoothed, running median color ridge lines, with the dotted
lines indicating one standard deviation; the ridges are computed using
stars with $F110W-F160W \leq 2.0$ only.  Objects with redder colors,
especially faint objects, are probably galaxies, consistent with the
roughly constant numbers of such objects from field to field.  For
reference, the horizontal dotted lines mark the location of the RGB
tip found from analysis of the LF histograms ($\S5.1$, Figure~7).

\begin{figure}
\vspace {-1.0in}
\epsscale{1.}
\plotone{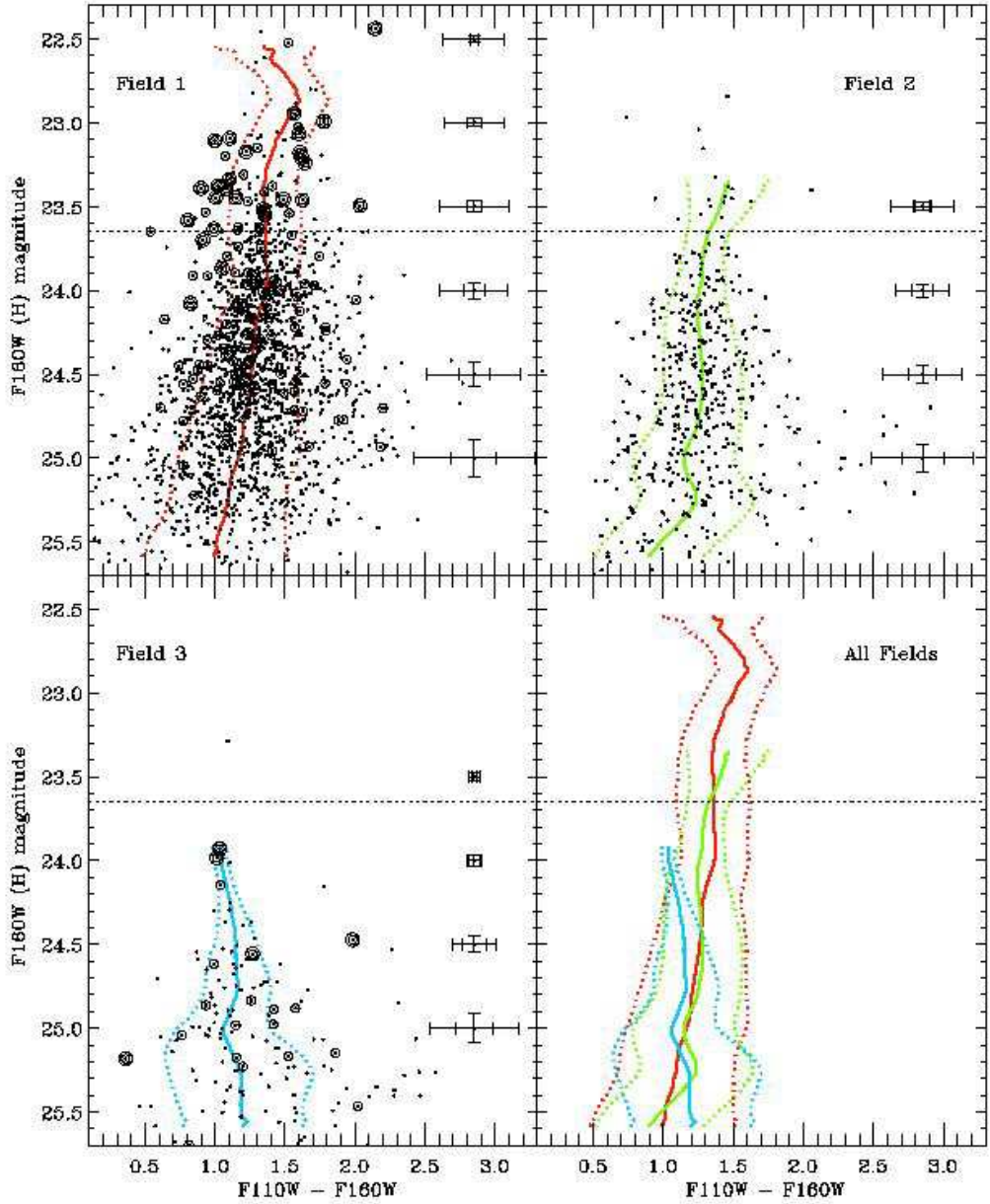}
\vspace {-0.5in}
\caption{\small Comparison of the color-magnitude diagrams of the three
NICMOS fields in NGC3379.  Circled stars have varied by 6.5 or 10
sigma (single and double circles, respectively; the second epoch data
for field 2 are not deep enough to confidently identify variables).
The mean (solid) and one standard deviation (dotted) ridge lines of
each field are shown and reproduced together for comparison in the
lower right panel; objects with colors $> 2.0$ are probably background
galaxies and have been clipped from the ridge line determinations.
There is no significant difference in the three populations spanning
nuclear distances of 9~kpc to 18~kpc.  The horizontal dotted line
marks the magnitude of the TRGB.  The sloping dashed line indicates
the 50\% completeness level for Field~1; confusion of the unresolved
stellar background is a greater limitation.}
\end{figure}

NGC3379 is at Galactic coordinates $l=133^\circ, b=57^\circ$.  There
are two obvious bright foreground stars in the images, one each in
Fields~1 and 3 (Figures 2 and 4).  Scaling from star counts in the
Hubble Deep Field (Williams et al.\ 1996; M\'endez \& Minniti 1999)
and infrared Subaru Deep Field (Maihara et al.\ 2001), we estimate
that over the magnitude range displayed in Figure~9, $22.3 < F160W <
25.7$, the expected number of Galactic foreground stars is zero.

Two sets of $F110W-F160W$ error bars are shown: the inner are the
median $F160W$ photometric errors reported by {\sc daophot},
determined in half-magnitude bins.  The outer error bars show the
$1\sigma$ widths of the $F110W-F160W$ color distribution at each
magnitude level, again determined in half-magnitude intervals.  The
bright limits of the ridge lines are determined solely by the extent
of the data and are not meant to suggest the locations of the RGB tip
or any other stellar evolutionary phase.  The three ridge and
dispersion lines are over-plotted for comparison in the lower right
panel of Figure~9.  There is no significant shift of the main RGB
locus from one field to another, and, given the small number of stars
in Field~3 and the similarity of the loci for Fields 1 and 2, these
data suggest that there are no substantial differences in the stellar
populations of the three fields.  Most of our subsequent analysis is
based on Field~1 alone, but by implication applies to the other fields
as well.

\subsection{Comparison to Theoretical Isochrones}

Determination of the metallicity and spread in abundance of the stars
in NGC3379 was a primary motivation of this project.  Using the
best-fit RGB tip magnitude of $F160W = 23.68$ found above, it is
possible to derive a distance to NGC3379, in turn allowing constraints
on its age and metallicity by comparing the NICMOS CMDs to theoretical
isochrones.  We have considered two independent sets of isochrones:
the latest version of the Bruzual \& Charlot (1993) models (BC; see
also Liu, Charlot \& Graham 2000), kindly provided in advance of
publication by S. Charlot, plus the isochrones of Girardi et al.\
(2002).  The BC isochrones are derived largely from the ``Geneva''
evolutionary tracks (Maeder \& Meynet 1989) while the Girardi set is
based on the ``Padova'' evolutionary tracks (Bertelli et al.\ 1994).

\subsubsection{Mean Metallicity and Age Constraints}

In Figure~10, upper RGB isochrones for 3, 5, 8, 10 and 15 (or 14) Gyr
populations with [Fe/H]=$-0.7$, 0.0, and $+0.4$ or $+0.2$ are
over-plotted on the ridge lines for Field~1.  Also plotted are the
standard deviation lines (dashed) of the observed NGC3379 giant
branch.  The isochrones have been shifted to match the observed RGB
tip magnitude, $F160W = 23.68$, in NGC3379.  The resulting individual distance
moduli are listed in the figure.  For the Girardi isochrones, the AGB
loci are also plotted (dotted lines).  The AGB in the BC
isochrones do not rise more than a few hundredths of a magnitude above
the RGB tip.

Both sets of isochrones demonstrate that the location of the RGB in an
infrared CMD depends primarily on metallicity while age is a second
order effect.  Neither set of [Fe/H]=$-0.7$ isochrones for any age
reproduces the mean color location of the RGB in NGC3379; the data
demand a much higher abundance.  The younger super-solar isochrones
are perhaps a reasonable match in principle, but the relatively bright
and blue turnoff of a majority 3-5~Gyr population would be
incompatible with the optical broad band colors and spectrophotometry
of NGC3379 (Terlevich \& Forbes 2002), although strictly, the
integrated light analysis applies only to the nuclear regions.  This
leaves open the somewhat unlikely possibility that the halo comprises
mainly very metal rich intermediate age stars.  Additional age
constraints based on the CMD come from consideration of the observed
AGB, which is better fit by the older ($>8$ Gyr) isochrones of Girardi
et al.\ (2002), as will be shown in $\S 6.1.4$.

\begin{figure}[th!]
\epsscale{1.0}
\plotone{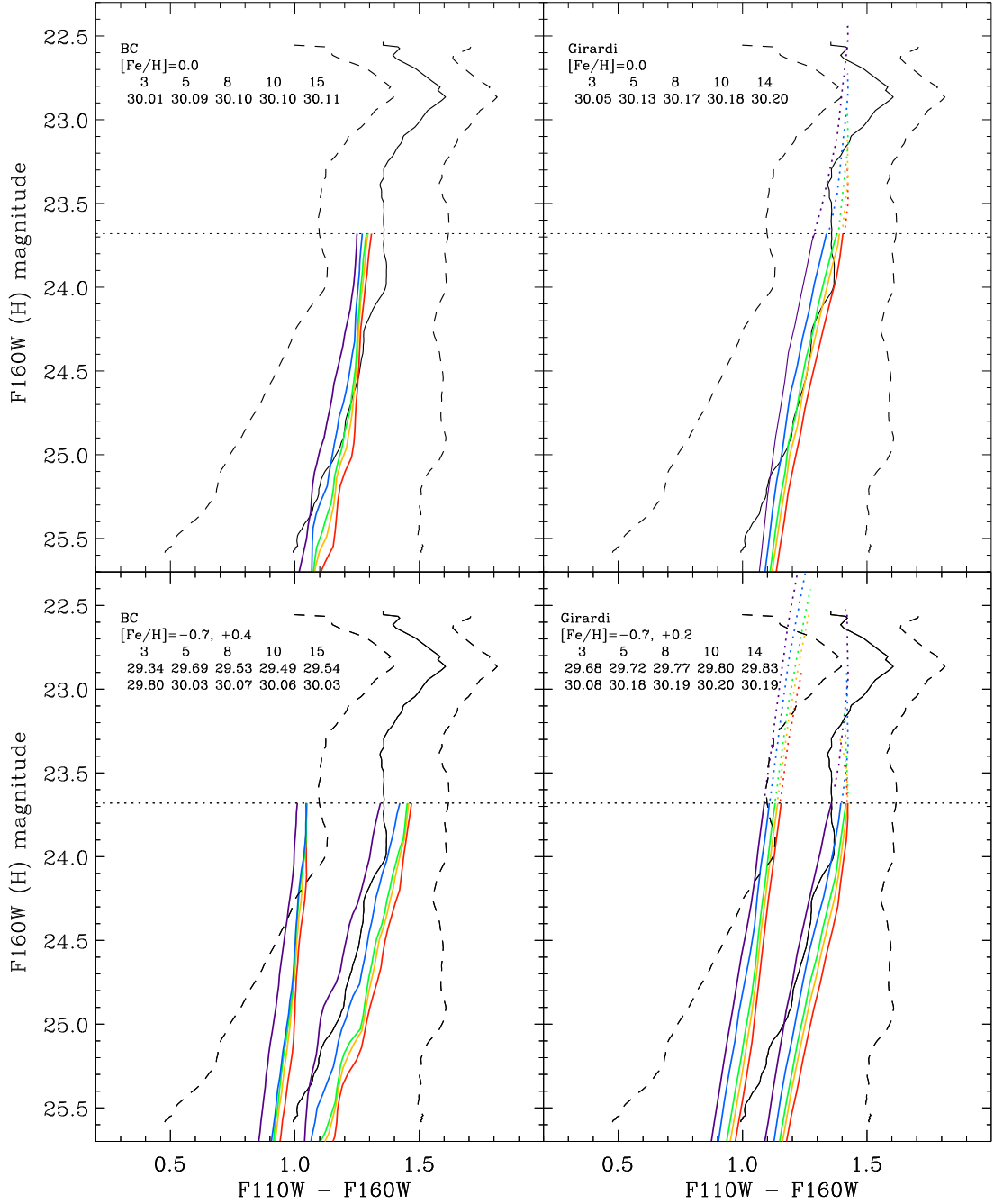}
\caption{\small Comparison of isochrone ages and abundances.  All isochrones
have been shifted so that their RGB tips are at $F160W = 23.68$, out
best-fit tip magnitude; the
corresponding distance moduli in each case are listed.  Both the BC
and the Girardi isochrones indicate a roughly solar mean abundance and
a mean age of 10 to 15~Gyr.  The dotted colored lines show the loci of the
upper AGB in the Girardi isochrones.}
\end{figure}

The solar abundance, 8, 10 and 14/15~Gyr isochrones of either set
bracket the RGB ridge line down to $F160W = 25$.  At this magnitude,
the observed RGB begins to get bluer.  
The blueward trend is from incompleteness in the $F110W$ band and
the onset of confusion close to the limit of the data (Appendix A).
The brightest 0.5 magnitude of the RGB of the solar BC isochrones do
not track the observed ridge line as well as the corresponding Girardi
isochrones.

The isochrones thus constrain the {\em mean} abundance of NGC3379 to
be close to solar metallicity, independent of the age of the stellar
population.  While the usual age-metallicity degeneracy still affects
the conclusions at some level, the {\em mean} age is likely to be in
the range $8-15$~Gyr, probably very close to 10~Gyr.  Given the
relatively small offsets in color due to age in the $F110W-F160W$
colors, and the relatively large observed width of the RGB, a
considerable spread in age above $\sim 8$~Gyr cannot be ruled out.

A metallicity close to solar is perhaps surprising so far out in the
halo, even of a giant elliptical.  One possible explanation for the
high abundance is that NICMOS is measuring the brightest, most metal
rich stars.  Looking at the CMDs and isochrones in Figures~9 and 11,
the metal poor populations are fainter and bluer.  Stars with
[Fe/H]$<-1.7$ will drop below the sensitivity range of the
observations.  The mean abundance of the {\em measured} stars is close
to solar, but a deeper census might reveal the existence of a bluer,
much lower abundance population with [Fe/H]$<-2$.  This effect is the
opposite of what happens in the optical where $V$ (and even $I$) band
RGBs become fainter with increasing metallicity, leaving just the
metal poor stars to be measured in shallow observations, resulting in
a lower mean abundance.  With the present NICMOS data we may be
overestimating the mean metallicity.  Complementary deep $V$ and $I$
band observations with the Advanced Camera for Surveys (ACS),
especially leveraged against these and/or additional NICMOS fields,
are required to document the full extent of the metallicity spread in
NGC3379.

\subsubsection{An Infrared RGB Tip Distance to NGC3379}

The sharpness of the break in the LF and the small photometric errors
at the RGB tip ($\pm 0.035$ magnitudes per star), provide a robust
location of the RGB tip from the maximum-likelihood analysis.
Adopting the solar metallicity, 10~Gyr isochrone as the best fit, the
implied distance modulus to NGC3379 is 30.10 (10.4~Mpc) for the BC
isochrone, $+0.08$ greater for the Girardi isochrone.  The major
contributors to the uncertainty in distance are the NICMOS photometric
calibration and the isochrone differences.  The NICMOS absolute
calibration is good to $\pm0.05$ magnitudes according to the NICMOS
website.  The Girardi et al.\ isochrones are systematically brighter
and redder than those of BC, the solar 10 Gyr RGB tip in particular by
0.08 magnitudes.  Much of this difference could arise from the
coarseness of the isochrone sampling of the RGB in steps of $\sim 0.05
- 0.07$ magnitudes.  Both the NICMOS calibration and the isochrone
differences are systematic in nature and we conservatively add them
linearly to derive a total uncertainty.  The total error in the
distance estimate is $\pm 0.14$ magnitudes, only 7\% in distance,
almost entirely from the uncertainties due to systematic differences
between the theoretical isochrones and the NICMOS absolute photometric
calibration, not the data.

Averaging of the two isochrone sets leads to $m-M=30.14 \pm 0.14$
(10.67~Mpc $\pm 0.74$) as the distance for NGC3379.  Our estimated
$1\sigma$ error includes the full covariance of the TRGB magnitude
with respect to the other parameters in the model, but does not
include possible systematic effects due to the abundance spread, which
we have not included in the maximum-likelihood analysis in $\S5.1$.
The mean abundance of NGC3379 is very near solar, and, fortuitously,
the solar abundance isochrones of either BC or Girardi et al.\ (2002)
have the most luminous (or nearly the most luminous) TRGB, so modest
contributions of stars from non-solar abundance populations will not
significantly affect the detection of the brightest (roughly solar
abundance) TRGB location.

The better match of the upper RGB and the better representation of the
brighter AGB stars by the Girardi isochrones supports a preference for
an increase in the distance to $m-M = 30.18$ (10.86~Mpc); a positive
choice of one set of isochrones over the other also removes a
contribution of 0.08 magnitudes to the systematic errors, which would
reduce the error estimate to $< 0.1$ magnitudes.  Regardless, the
derived distance is in good agreement with estimates by other
techniques (Gregg 1997; Hjorth \& Tanvir 1997), and consistent at the
$1\sigma$ with the results of Sakai et al.\ (1997) who found
$m-M=30.30\pm0.14$ (random errors) $\pm 0.23$ (systematic) from the
$I$-band luminosity of the RGB tip using WFPC2 imaging.

\subsubsection{Metallicity Spread}

\begin{figure}[t!]
\epsscale{1.0}
\plotone{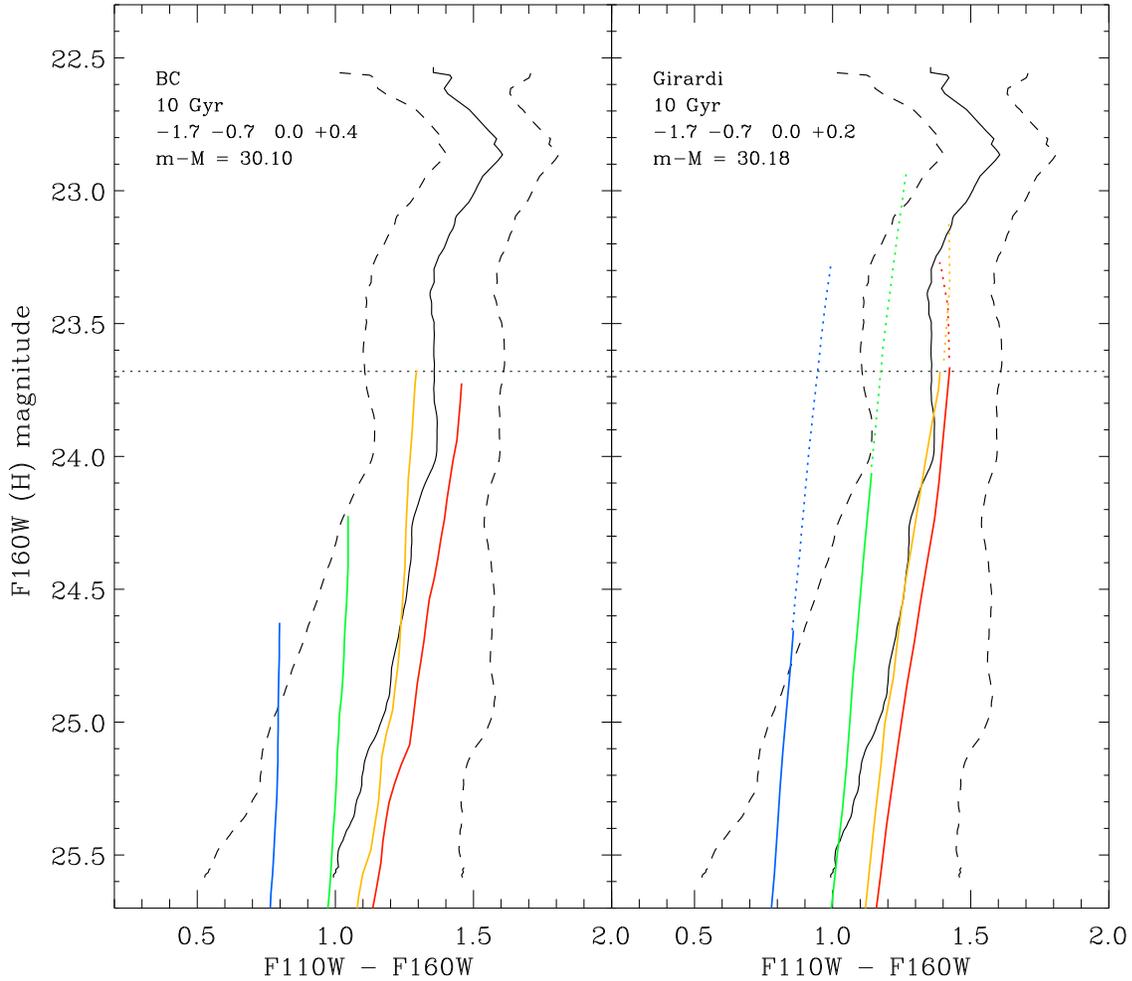}
\caption{\small The 10 Gyr isochrones of various metallicities, all shifted
to the same distance modulus (30.10 for the BC isochrones, 30.18 for
the Girardi), compared to the mean
color-magnitude distribution (solid black line) of NGC3379.  The
observed RGB tip magnitude is indicated by the horizontal dotted
line.  The dotted colored lines show the upper AGB loci of the
Girardi isochrones.  The one standard deviation dashed ridge lines
are consistent with the presence of stars with the wide range of
metallicities shown. }
\end{figure}

The metal poor isochrones are not a good match to the mean RGB of
NGC3379, but provide constraints on the metallicity spread.  From
examining the isochrone sets in Figure~10, it is clear that it is
impossible to account for the RGB width solely with an age spread;
most or all of the intrinsic width must be due to a spread in
abundance.  In Figure~11, over-plotted on the Field~1 ridge and
dispersion lines are BC and Girardi 10~Gyr isochrones for a range of
abundance, all shifted to the same distance modulus, 30.10 for BC and
30.18 for Girardi.  In both isochrone sets, the lower abundance RGB
tips have lower luminosity and are shifted to bluer colors, naturally
tracing the blue envelope of the observed RGB stars.  A range of
metallicity will thus increase the intrinsic (and observed) width of
the RGB at fainter magnitudes, even without a contribution from
increasing photometric errors, because there simply are no extremely
metal-poor RGB stars brighter than $F160W\approx 24.5$.  The
super-solar BC isochrone also has a somewhat fainter RGB tip
luminosity, perhaps signaling the same effect on the high metallicity
side.

Consequently, the isochrones predict that for a heterogeneous
abundance stellar population, the {\em spread} in metallicity will be
a function of magnitude over the upper $\sim 1.5$ magnitudes of the
RGB.  The color distributions in four 0.5 magnitude intervals
beginning at the RGB tip are plotted in Figure~12.  Over-plotted on
each is a best-fit gaussian with the same standard deviation and
median, normalized to the total number of stars.  The upper abscissa
is an approximate metallicity scale computed simply by dividing the
isochrone metallicity range (2.1 dex) by the color difference at the
central $F160W$ in each interval.  This is approximate and plotted
merely for reference; the change in color is not strictly a linear
function of abundance, nor constant as one goes down the RGB.

\begin{figure}[ht!]
\vspace {-0.45in}
\epsscale{0.85}
\plotone{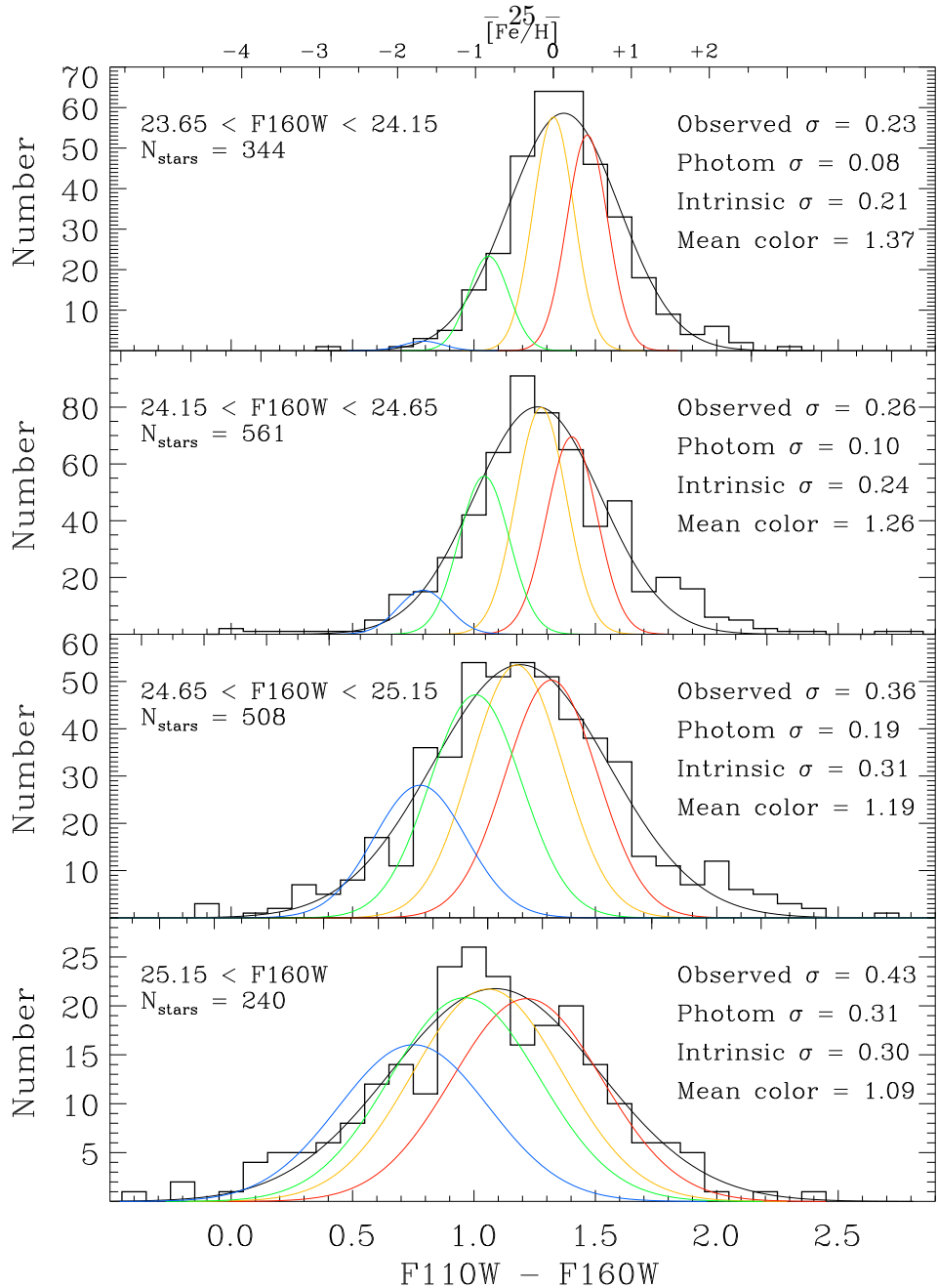}
\caption{\small Observed color spread of the RGB (histograms) in four
magnitude intervals compared to the widths expected from photometric
errors alone, as determined from artificial star tests.  The broad
gaussian in each panel is a robust fit to each histogram; the narrower
histograms show the expected widths and locations for hypothetical
single abundance populations with the metallicities shown in
Figure~10, their peaks scaled to fit on the broad gaussian.  These
qualitatively show the relative contributions of each abundance
interval, mainly illustrating the increasing contribution of the metal
poor stars at fainter magnitudes.  We have used the BC isochrones for
this analysis because of their higher maximum Z.}
\end{figure}

For comparison, we also plot gaussians at the mean color location of
the 10~Gyr BC isochrones with [Fe/H] = $-1.7, -0.7, 0.0, +0.4$.  Each
has a $1\sigma$ width equal to the spread in color caused by crowding
and photometric errors, as determined by our artificial star tests
(Appendix A).  The peaks are simply scaled to the local value of the
total gaussian fit and so these are meant to be suggestive only; the
abundance distribution of the stellar population is, as far as we can
tell from our data, continuous.  The near absence of any contribution
from extremely metal-poor blue stars in the brightest interval is
evident, as discussed above.  Such blue stars become
numerous at magnitudes $> 24.5$, once the tip of the fainter
metal-poor RGB enters the plots.  The presence of stars redder than
the [Fe/H]$=+0.4$ isochrone implies a very metal-rich population.  The
intrinsic width of the color distributions increases gradually from
0.21 to 0.31 magnitudes over the three brightest intervals, but does
not increase further in the faintest, suggesting that we may be seeing
the full extent of the abundance spread, but deeper data are required
to confirm this.  The faintest interval is essentially confusion
limited (see Appendix A), however, so while it is consistent with the
next brighter interval, it cannot bear much weight in the analysis.

Using the color-to-metallicity conversions computed from the
isochrones, the metallicity spreads in each 0.5 magnitude interval on
the RGB are $\pm 0.6$, $\pm 0.8$, $\pm 1.2$, and $\pm 1.2$~dex.  The
tail to very red objects, $F110W-F160W\gtrsim2.$ is perhaps due to
unresolved galaxies.  There is a suggestion of a break in the middle
two color histograms at $F110W - F160W \approx 1.65$, perhaps an
indication of the transition between the stellar population and the
background galaxies.  The blueward trend of the most metal-rich
isochrones with fainter magnitudes combined with photometric errors
may produce this break on the red side.  Extrapolating from the BC
isochrones, we estimate that this color corresponds to [Fe/H]$ \approx
+0.8$.  Stars this metal-rich have been shown to exist in the inner
Milky Way halo (Rich 1986; Minniti et al.\ 1995; Sadler, Rich, \&
Terndrup 1996), coexisting with extremely metal poor objects of
$[Fe/H] = -2$ (Blanco 1984), similar to the situation we find in
NGC3379.

\subsubsection{The Asymptotic Giant Branch Population}

The characteristics of the AGB population evolve with age and are a
function of metallicity, so in principle the AGB can be a stellar
populations diagnostic.  To be useful, though, the AGB contribution
must somehow be separated from the RGB, which is problematic when
photometric errors and stellar physics cause them to completely
overlap fainter than the RGB tip.

\begin{figure}[t!]
\plotone{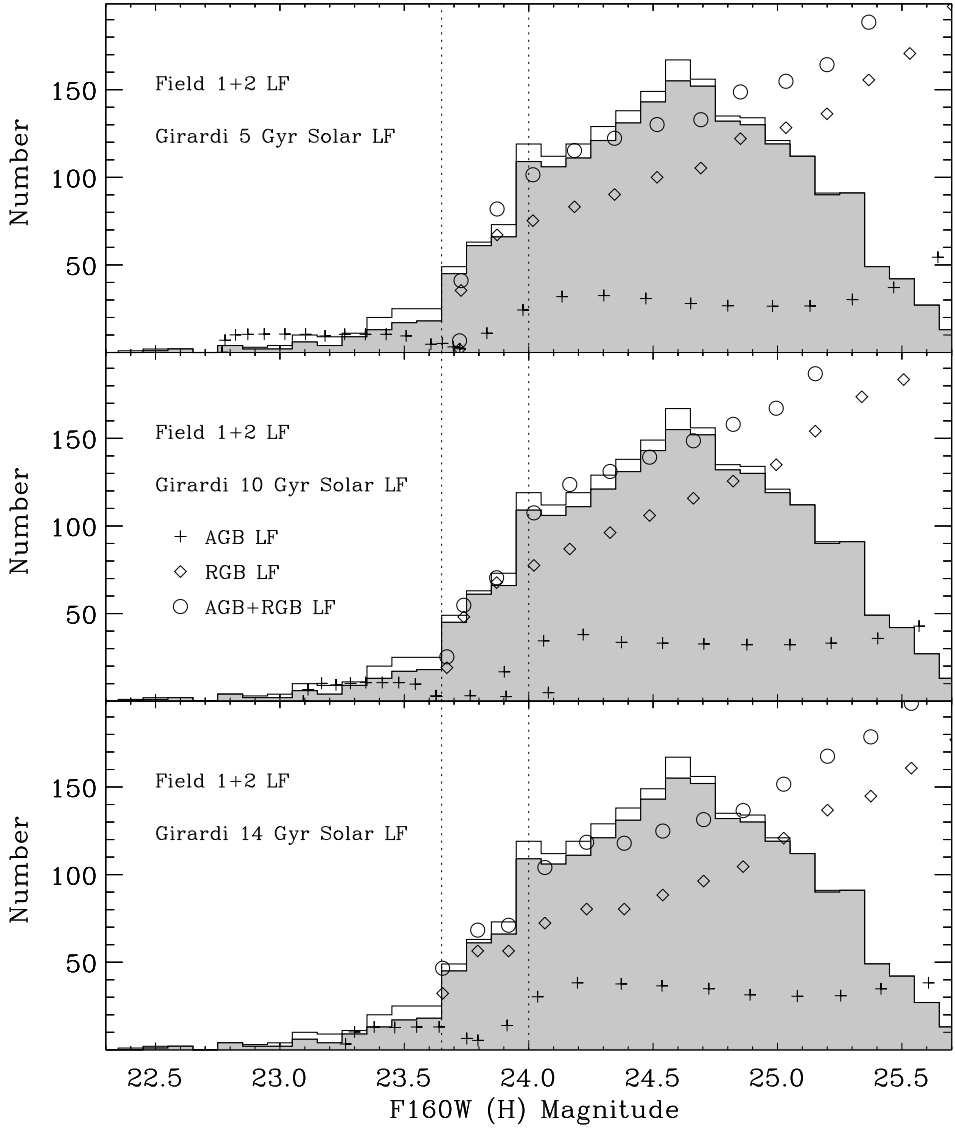}
\caption{\small Theoretical luminosity functions (points) based on the
Girardi solar metallicity isochrones were computed using a Salpeter
initial mass function (x = 1.35) for comparison to the observed LF,
with and without variables (histograms), of NGC3379.  The 10 and 14
Gyr models, or a composite, can account for the majority of AGB stars
above the RGB tip.  The break in the observed LF at $F160W = 24.0$ is
qualitatively accounted for by a decline in the brighter AGB
population. }
\end{figure}

Stars significantly brighter than the RGB tip, however, can be
relatively easy to isolate; work by Mould \& Aaronson (1979) showed
that such AGB stars mark the presence of intermediate age populations,
$ \lesssim 3$~Gyr.  Yet the existence of stars up to $\sim 1$
magnitude brighter than the RGB tip in metal rich Galactic globular
clusters (Guarnieri, Renzini, \& Ortolani 1997), advises caution in
interpreting such stars as an unambiguous signal of younger
populations.  The behavior of the AGB in the Girardi isochrones
(Figures~10, 11) also shows that stars brighter than the RGB tip by up
to $\sim 1$ magnitude can be expected even in ancient populations.
We find 98 stars above the tip of the RGB in Field~1 (Table~3,
Figure~9), 38\% of which are variable, and an additional 19 bright
stars in Field~2, plus one more in Field~3.  All but 13 of these are
within 0.6 magnitudes of the RGB tip.  

To test the conclusions from $\S 6.1.1$ that the population of NGC3379
is old and solar metallicity, we computed theoretical LFs using a
Salpeter initial mass function (IMF) using the Girardi 5, 10, and 14
Gyr solar isochrones.  The resulting comparison with the observed
$F160W$ LF is necessarily approximate as the isochrones are sampled in
uneven mass bins determined by theoretical considerations while the
observed LF has been computed in equal magnitude bins.  We have not
interpolated the isochrones because this smooths out sharp features
such as the RGB tip and also because the isochrones are not monotonic
functions in luminosity in regions of interest, making interpolation
problematic.  Fortuitously, the isochrone mass intervals correspond to
magnitude intervals of 0.1 to 0.15 along the RGB.  We have, however,
rebinned the theoretical AGB LF to the same magnitude intervals as the
theoretical RGB to permit easy summing and have scaled the total LFs
by eye to give an approximate match to the observed LF (Figure~13).

The 10 or 14 Gyr isochrones, or a composite, can account for the AGB
stars up to 0.6 magnitudes brighter than the RGB tip.  These stars
then do not require an intermediate age or younger population in
NGC3379.  Any claim of such a population would rest on the 13 yet
brighter AGB stars ($F160W < 23.0$).  Such stars are predicted by the
5~Gyr isochrone, but the numbers of observed AGB stars above the TRGB
limit any 5~Gyr contribution to $\sim 20\%$ of the total.  The
brightest AGB stars are more naturally explained by a contribution
from the substantial [Fe/H]$=-0.7$ component needed to account for the
metallicity spread (c.f.\ Figure~11); this population will generate
AGB stars up to 0.8 magnitudes above the RGB tip.  Blends of stars in
the images, or even evolving blue straggler or mass transfer binaries
are also viable explanations for these few extremely bright AGB stars.
We conclude that there is no compelling evidence for any substantial
population with an age $< 10$~Gyr.

An explanation for the second LF break at $F160W=24$ also emerges: it
is the top of the stably evolving AGB.  Between $F160W=24$ and the RGB
tip, the AGB stars in the Girardi isochrones evolve rapidly, thinning
out the AGB LF, and it is in this region that they exhibit
non-monotonic luminosity behavior.  The 10 and 14 Gyr ages reproduce
this feature somewhat better than the 5 Gyr model, suggesting that the
structure of the LF may be useful in constraining the age of the
observed population.  Detailed comparisons require isochrones more
finely sampled in mass (luminosity) and perhaps also better statistics
in the observed stars.

\subsection{Comparison to Other Stellar Systems}

Comparison of the NICMOS data for NGC3379 to the theoretical
isochrones of BC and Girardi has resulted in a consistent, perhaps
even plausible, picture of the stellar populations in the halo of a
giant elliptical.  We next compare the data to similar observations
of various well-studied stellar systems.

\subsubsection{Galactic Globular Clusters and the Bulge}

The stellar population of Galactic globular clusters is simple, well
understood, and has much IR data available.  Though the Milky Way
Bulge population is less well-determined, high quality IR data exists
and provides another point of comparison for a system which is perhaps
more similar to NGC3379.  Comparison with the extensive Milky Way
ground-based data requires color transformations from the CIT $J \& H$
bands to NICMOS $F110W \& F160W$.  We found it necessary to derive our
own transformation relation for this purpose; details are given in
Appendix~B.

\begin{figure}[t!]
\epsscale{1.0}
\plotone{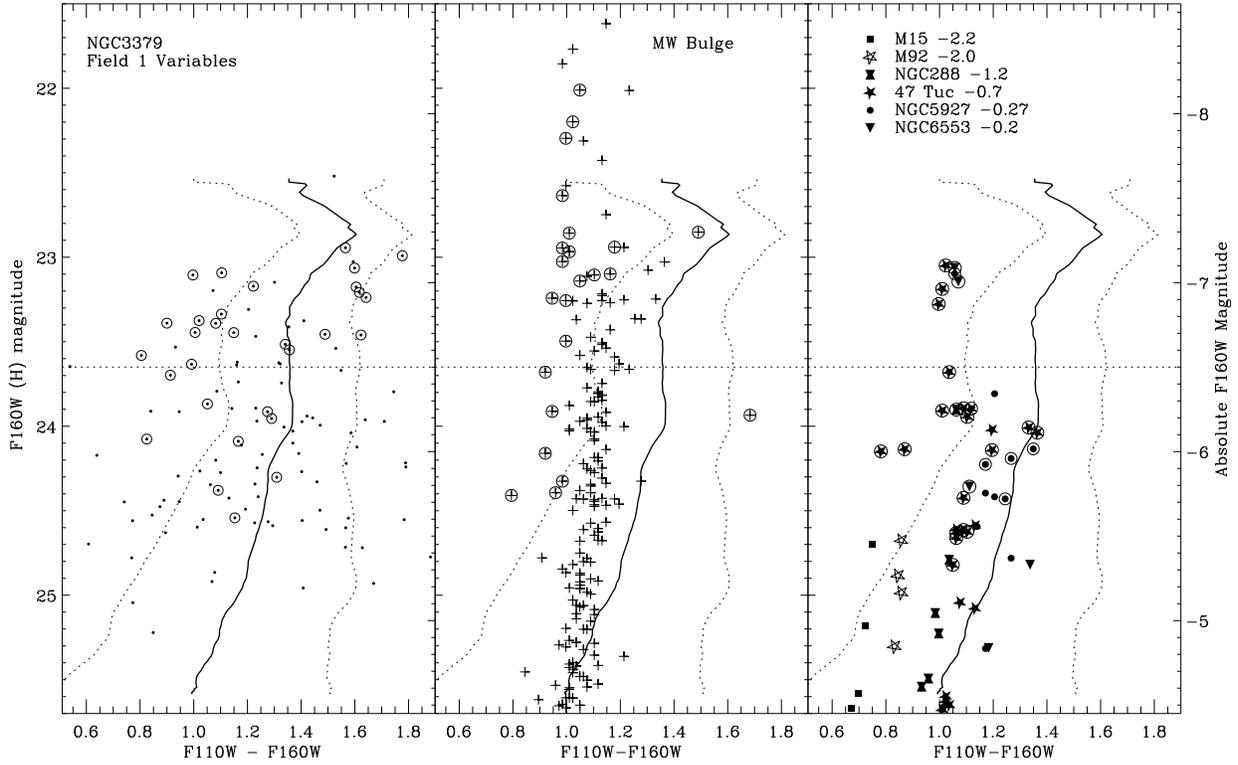}
\caption{\small Comparison of NGC3379 RGB ridge line and variables for
Field~1 with the Milky Way Bulge and globular clusters with a range of
abundances.  Galactic data have been shifted to the distance of NGC3379.}
\end{figure}

In Figure~14, we compare the transformed ground-based CIT $J$ and $H$
photometry of the Milky Way Bulge (Frogel et al.\ 1990; Tiede et al.\
1995) and several representative Galactic globular clusters (Frogel et
al.\ 1981, 1983; Davidge \& Simons 1994) to the Field~1 NGC3379 RGB/AGB
ridge line and variable star distribution.  The data have been
shifted to the distance modulus of NGC3379; individual distance and
reddening estimates are taken from the photometry sources.  The
globulars range from [Fe/H]$ = -2.2$ (M92) to $-0.2$ (NGC6553).  Known
Bulge and globular variables are circled.

Given the wide spread in abundance known to exist in the MW Bulge
(Rich 1986; Sadler et al.\ 1996; Minniti et al.\ 1995; Blanco 1984),
it is puzzling that its RGB/AGB locus is so tight, implying a
relatively narrow range of metallicity.  Comparison with Figure~10
reveals that the older [Fe/H]$ = -0.7$ isochrones are a good match for
the bulge giants, similar to 47~Tuc, consistent with the analysis of
(Frogel et al.\ 1990).  The Bulge clearly has AGB stars much brighter
than those in NGC3379, up to 2 magnitudes brighter than the RGB tip;
in fact, we have omitted a number of Bulge stars still brighter than
those plotted here.  Frogel et al.\ (1990) concluded that no
intermediate age stars were necessary to account for the bright AGB
population of the Bulge.  The Girardi isochrones, however, indicate
that an intermediate age ($\lesssim 5$~Gyr) population is required to
account for the AGB stars more than 1 magnitude above the RGB tip,
which exist in some of the Bulge fields.  It is not clear from the
literature accounts whether blending of stars in the crowded fields or
perhaps a large spread in distance can account for all of these very
bright AGB stars.

The collection of Galactic globular clusters is a better approximation
to the data for NGC3379.  The width of the RGB, the distribution of
variables, and the luminosity of stars brighter than the RGB tip all
point to a purely old population in NGC3379 with a wide metallicity
spread, consistent with the analysis based on theoretical isochrones
above.  The variables up to 1 magnitude brighter than the RGB tip in
NGC3379 attest to its having a relatively metal rich component,
similar to 47~Tuc or greater, as such stars are not found in the metal
poor globulars with [Fe/H]$ < -1$ (Frogel 1983).  The lack of yet
brighter AGB stars in our NICMOS data amounts to lack of evidence for
a population $\lesssim 5$ Gyr, as seen in regions known to harbor
intermediate age populations, such as the Magellanic Clouds (Frogel et
al.\ 1990).  

\subsubsection{NGC5128}

The peculiar elliptical NGC5128 hosts an active galactic nucleus
(Soria et al.\ 1996) and powerful radio source (Centaurus~A).  Its
prominent dust lane and stellar and gaseous shells are only part of
the strong evidence of recent and perhaps multiple merger events,
possibly the trigger for the nuclear activity and radio emission.
NICMOS observations in F110W and F160W have been carried out by
Marleau et al.\ (2000) for a field in the halo of NGC5128,
$8\arcmin50\arcsec$ (9~kpc) from its nucleus, reaching slightly
greater effective depth relative to the RGB as our data for NGC3379.

\begin{figure}[t!]
\vspace {-0.5in}
\epsscale{0.8}
\plotone{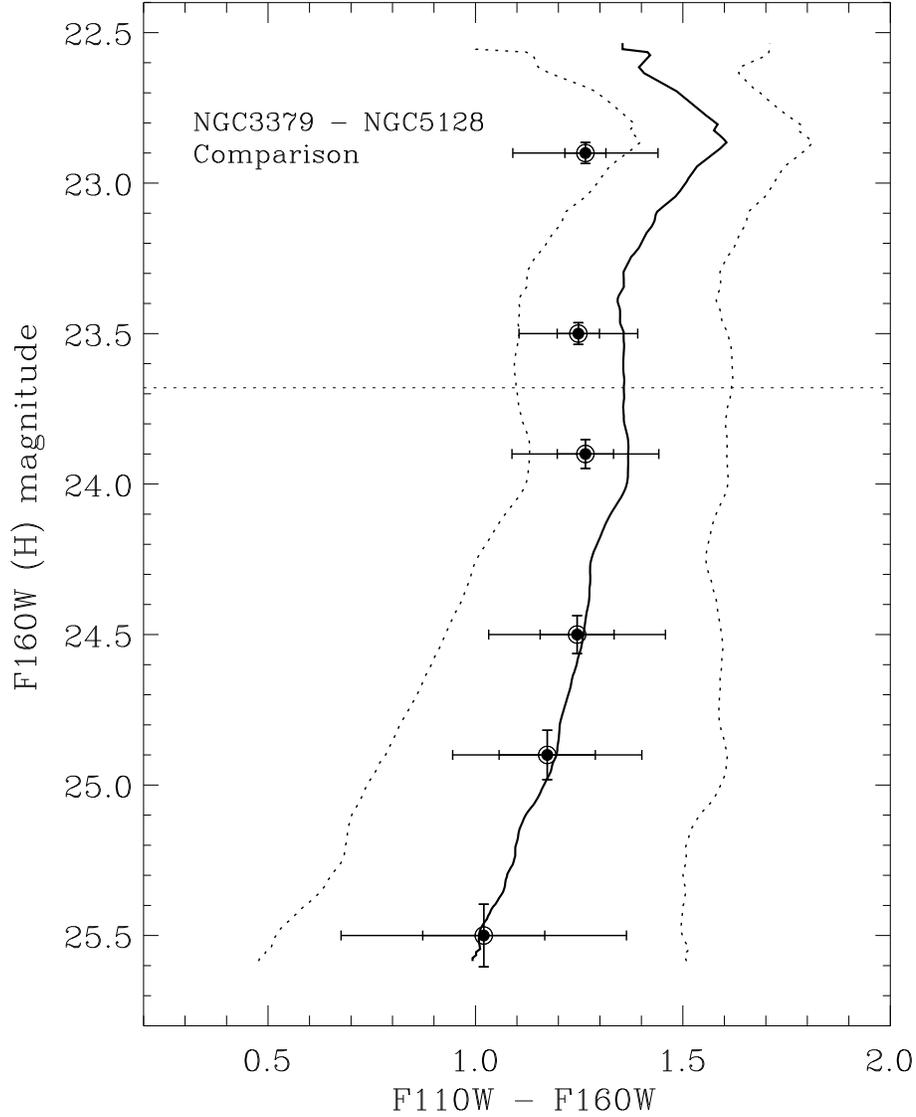}
\caption{\small NGC3379 Field 1 CMD ridge line and dispersion (solid and
dotted lines) compared with that of NGC5128 (points and error bars) as
measured by Marleau et al.\ (2000).  The NGC5128 points have been
shifted from the TRGB distance modulus of 27.98 found by Harris et
al.\ (1999) to our NICMOS TRGB distance modulus for NGC3379.}
\end{figure}

In Figure~15 we plot both the NGC3379 Field~1 RGB/AGB ridge lines and
the mean $F110W-F160W$ color locations from Marleau et al.\ (2000) for
NGC5128, shifted from the TRGB distance modulus of NGC5128
($m-M=27.98$, Harris, Harris, \& Poole 1999) to our derived NICMOS
TRGB distance modulus for NGC3379.  The photometric errors of the two
data sets are comparable at a given RGB luminosity.  Despite the
uncertainties in extinction corrections and distances to the two
objects, plus the independent and different natures of the reductions
and analysis -- they used an empirical PSF and did not employ
drizzling -- the mean loci of the lower RGB points differ in color by
only $\sim 0.02$ magnitudes.  This is reassuring evidence that there
are no serious systematic calibration differences between the studies.

Two differences in the populations of the galaxies emerge, however.
The observed color spread of the NGC3379 RGB is considerably greater,
$\sigma = 0.2$ magnitudes for NGC5128 compared to $\sigma = 0.4$
magnitudes for NGC3379 over the fainter half of the data; this cannot
be accounted for entirely by our slightly greater observational
errors.  It is most easily understood as an abundance range only half
as great in NGC5128.  Second, the brighter NGC5128 RGB points lie to
the blue of the ridge line of NGC3379.  Marleau et al.\ argue for a
significant contribution from an intermediate age ($1-5$~Gyr)
population, based on the presence of numerous stars, probably
belonging to the AGB, up to almost 2 magnitudes brighter and several
tenths bluer than the RGB tip, much like those of the Milky Way Bulge.
The stars above the RGB tip in NGC3379 are concentrated within 0.6
magnitudes of the tip, completely consistent with older, metal rich
stars.  An increased population of these brighter and bluer AGB stars
from a younger component in NGC5128 can easily account for the 0.1 to
0.25 magnitude shift of its upper RGB/AGB locus relative to NGC3379.
The number of bright AGB stars seen in NGC5128 is $\sim 10-20$, so the
contrast with NGC3379 is probably not simply small number statistics,
underscoring the conclusion that NGC3379 has a significantly smaller,
if any, intermediate age population.  Such a difference is certainly
consistent, perhaps expected, from the respective morphologies,
NGC3379 the quintessential normal elliptical and NGC5128 being quite
peculiar.

The coincidence of the NGC3379 and NGC5128 RGB loci, along with our
recalibration of the ground-based IR to NICMOS photometry
transformation, implies that the mean abundance of NGC5128 in the
field studied by Marleau et al.\ (2000) should be revised upward from
their [Fe/H]=$-0.76$ to near solar.  In support of this, Harris et
al.\ (1999), using $V$ and $I-$band WFPC2 data find that the peak of
the abundance distribution for halo stars in NGC5128 is at
[Fe/H]$=-0.3$ in a field twice as far from the nucleus, where one
might expect the abundance to be lower than in the Marleau et al.\
field.  Having suffered a recent merger, NGC5128 is a complex system
probably not in equilibrium nor well-mixed, so perhaps such
field-to-field differences are real.  Harris et al.\ (1999) also find
a large abundance spread in NGC5128, with stars at least as metal poor
as [Fe/H]$=-2$ up to at least [Fe/H]$=+0.2$, not unlike what we find
for NGC3379 (Figure~12).  Their photometry reaches about 1 magnitude
farther down the RGB, allowing them to discern that the metallicity
distribution of NGC5128 has two components.  Such a conclusion is not
ruled out for NGC3379 by our data, but a deeper exploration of its
giant branch is needed to explore the similarities with NGC5128 in
greater detail.

Rejkuba, Minniti, \& Silva (2003) have found large numbers of long
period variables (LPV) in NGC5128, with periods ranging from 150d to $>
800$.  The existence of variables with P$ > 300$d is consistent with
NGC5128 having a significant intermediate age population.  Variables
with periods greater than several hundred days have ages $1-5$ Gyr,
whereas AGB variables with periods of $< 250$ days are much older
(Frogel 1983; Hughes \& Wood 1990).  With just two epochs, we cannot
constrain the periods of the variables in NGC3379; however, the
necessary observations are well within reach of either the refurbished
NICMOS camera or ground-based IR cameras with adaptive optics.  Period
determinations for the LPV stars in NGC3379 is perhaps the most
sensitive test for detecting and quantifying any intermediate age
population in this normal elliptical galaxy.

\section{A Note on the Morphological Type of NGC3379}

There has been extensive discussion in the literature of the true
morphological type of NGC3379 (e.g., Van den Bergh 1989; Capaccioli et
al.\ 1991; Statler 1994).  Is it a bona fide normal elliptical galaxy
or is it a face-on S0?  On one hand, this is an extremely important
issue, for the present picture of the origin and evolution of
elliptical galaxies is significantly different from that supposed for
S0s (Gregg 1989), thus the morphology of NGC3379 has ramifications for
the conclusions we can draw for galaxies in general.  On the other
hand, the question of its morphological type is moot: with its
standard colors, spectrum, and appearance, it might as well be
considered a prototypical elliptical.  If after such detailed
investigations, we are unable to discern the morphological type of
NGC3379, at a distance of only 10 Mpc, then it is practically
impossible to establish the ``true'' morphology of other early type
galaxies at greater distances in clusters such as Coma, let alone at
high redshift.  There will be many NGC3379 clones in these
more-distant samples, so the population of NGC3379 is bound to be
representative of an appreciable fraction galaxies taken to be early
type, fundamental plane objects.

\section{Conclusions}

Our conclusions are that NGC3379 has:

\begin{itemize}

\item a distance of $10.8\pm0.7$~Mpc, derived from comparison of the
  tip of the RGB in the F160W ($H$) band to the theoretical isochrones
  of Bruzual \& Charlot (1993) and Girardi et al.\ (2002);

\item a mean metallicity of roughly solar, perhaps slightly less, 

\item a large abundance spread with [Fe/H] ranging from $-2$ to +0.8;

\item no significant change in the mean abundance over almost a factor
  of two in distance from the nucleus,  9~kpc to 18~kpc.

\item a relatively old mean age, in the range 8 to 15~Gyr;

\item no significant intermediate age population $\lesssim 5$ Gyr, but
  a large age spread from 8 to 15 Gyr cannot be ruled out

\item a large population of bright AGB and RGB variables, similar to
  old stellar populations in the Galactic Bulge and globular
  clusters.

\end{itemize}

The overall stellar population of NGC3379 resembles a collection of
old Galactic globular clusters having a wide distribution of
metallicities, but including a significant contribution from
super-solar abundance stars.  If the large abundance spread and the
significant population of metal poor stars extends to the central
regions of NGC3379, then this must be taken into account when modeling
and interpreting the integrated spectrum, where the infamous
age-metallicity degeneracy problem is acute.  This investigation has
merely scratched the surface of what is now possible to glean from
detailed, star-by-star analyses of stellar populations at this
distance.  Exploring the metallicity range in more detail, especially
the metal poor stellar population, awaits deep optical imaging with
the Advanced Camera for Surveys on HST.  Determining the periods and
period distribution of the luminous IR variable stars is the most
sensitive way to better constrain any intermediate age population, and
this can be done with NICMOS or, eventually, with ground-based
adaptive optics observations.  Both optical and IR investigations can
build on the work described here, refining our understanding of the
formation and evolution of early type galaxies.

\acknowledgments

Support for this work was provided by NASA through grant number
GO-7878 from the Space Telescope Science Institute, which is operated
by AURA, Inc., under NASA contract NAS5-26555.  Part of the work
reported here was done at the Institute of Geophysics and Planetary
Physics, under the auspices of the U.S. Department of Energy by
Lawrence Livermore National Laboratory under contract
No.~W-7405-Eng-48.
DM is supported by FONDAP Center for Astrophysics 15010003.
We thank Stephane Charlot and Leo Girardi for supplying isochrones in
the NICMOS filters and also for helpful discussions.

\pagebreak

\section{Appendix A: Artificial Star Tests}

Our three target fields were selected to span a large range of surface
brightness, with the innermost still having low enough stellar density
so as not to be crippled by confusion of sources as bright as the RGB
tip, following the analysis and prescription laid out by Renzini
(1998).  To assess the impact of crowding on our data, we used the
{\sc daophot} routine {\sc addstar} to implant artificial stars in the
reduced images of Field~1.  In each trial, 100 stars of a given
$F160W$ magnitude and having a color of 1.1 -- about $1\sigma$ to the
blue of the mean RGB color -- were inserted on a $10\times10$ grid,
roughly evenly spaced over the image.  Fifty sets of $F110W$, $F160W$
image pairs were made for each input $F160W$ magnitude, ranging from
22.75 to 26.5 in steps of 0.25 magnitudes.  The locations of the
artificial stars were randomly perturbed within a 20 pixel radius for
each run, but held fixed within each $F110W, F160W$ image pair.  Each
of the 13 magnitude intervals then has 5000 artificial stars of known
input brightness and color.  A second and equally extensive trial was
made with stars having a color of 1.5, about $1\sigma$ to the red of
the mean RGB.

The same reduction procedure described above for the real data was
used to find and measure all the stars in each artificial star image;
and then compared to the input artificial star lists.  Table~4
summarizes the the input star magnitudes and the means and dispersions
of the recovered magnitudes and colors.  These can be compared to the
photometric errors reported by {\sc daophot} (Table~2).

From these tests, we can conclude the following: At F160W=24.68, one
magnitude below the RGB tip, we are 80\% complete.  The mean recovered
$F160W$ magnitudes at this level are too bright by $0.12$ magnitudes
because of faint stars blending to increase the measured brightnesses.
The mean recovered colors of the inserted $F160W=24.68$ stars are
somewhat less affected: those with $F110W-F160W = 1.1$ ($\sim 0.2$
magnitudes bluer than the mean RGB), are on average measured to be too
red by 0.05, while those with $F110W-F160W = 1.5$ ($\sim 0.2$
magnitudes redder than the mean RGB) are found to be too blue on
average by 0.07 magnitudes.  At the RGB tip itself, the systematic
error due to crowding is only 0.01 magnitudes in brightness or color,
thus crowding has no significant effect on the distance estimate.  At
faint magnitudes, the mean recovered color of both the red and blue
star tests converge at $\sim 1.28$, which we take to be the color of
the unresolved faint star background.

Fainter than an input magnitude of $F160W = 25$, the recovered
magnitudes saturate at $\sim 24.7$, signaling the confusion
limited reliability level of the data.  

\begin{figure}[t!]
\epsscale{1.}
\plotone{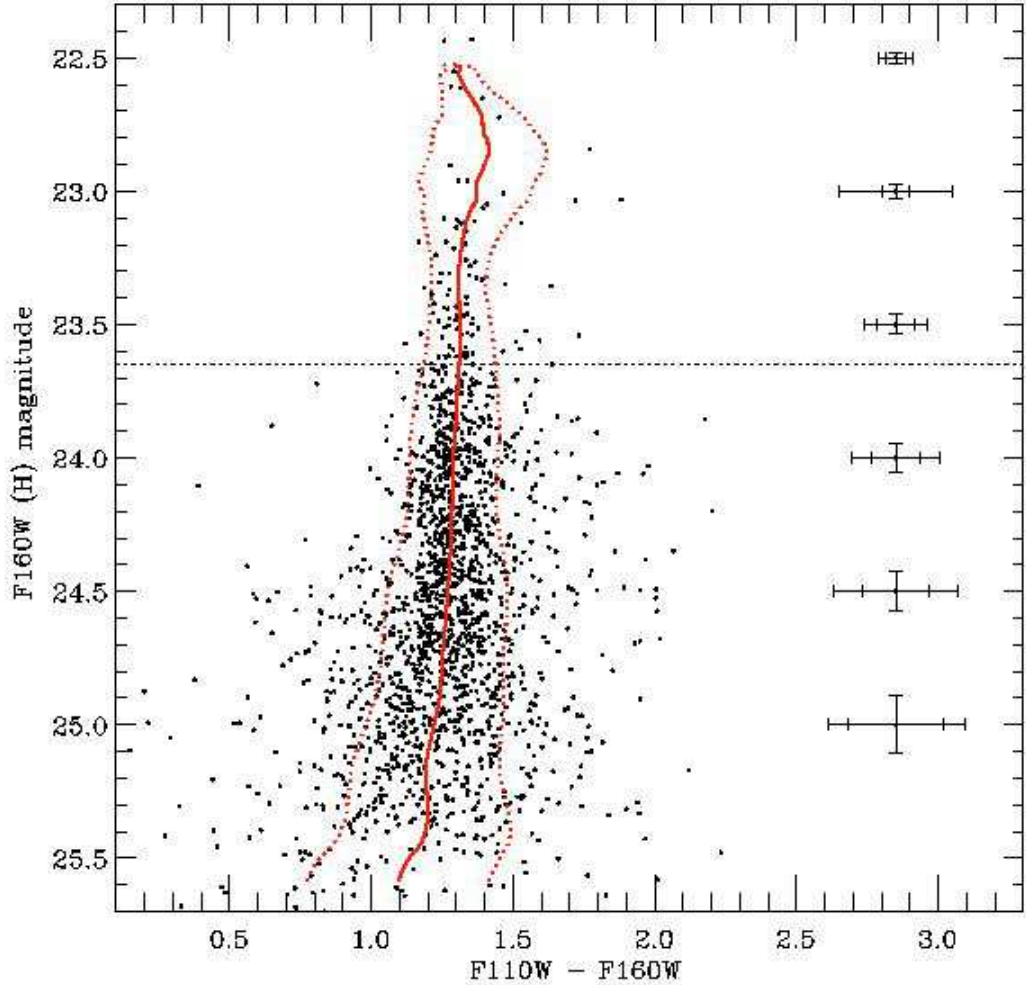}
\caption{\small A1: Color magnitude diagram for artificial star images
where all stars have $F110W-F160W = 1.30$.  The recovered RGB is
widened by crowding over purely photometric errors, but is still
significantly narrower than the observed spread (Figure~8).
}
\end{figure}

The spread in color of the recovered artificial stars is nearly as
great as the observed spread, but this results from the combined
effects of crowding and the intrinsic color spread of the real stars.
To determine the amount of spread in the RGB directly caused by
crowding, we created two new artificial images.  Beginning with the
residual noise images left after PSF subtraction by {\sc allstar}, we
added back artificial stars at the same locations as all the detected
stars; for the $F160W$ image, we used the detected magnitudes,
effectively reconstructing the original $F160W$ image, but with a
little extra Poisson noise which {\sc addstar} includes.  For the
$F110W$ image, we added stars at all the locations of the detected
stars, but adjusted their magnitudes so that {\em every} star had the
same color, $F110W-F160W = 1.3$, about the mean of the RGB.  Running
the {\sc daophot} suite on these images in a manner identical to that
used on the real data produces the CMD in Figure~A1; this figure
should be compared to the real data CMD of Figure~9.  The recovered
spread of the artificial RGB is greater than the median photometric
errors, but is quite a bit less than the observed RGB width of
NGC3379.

\begin{figure}[t!]
\epsscale{0.85}
\plotone{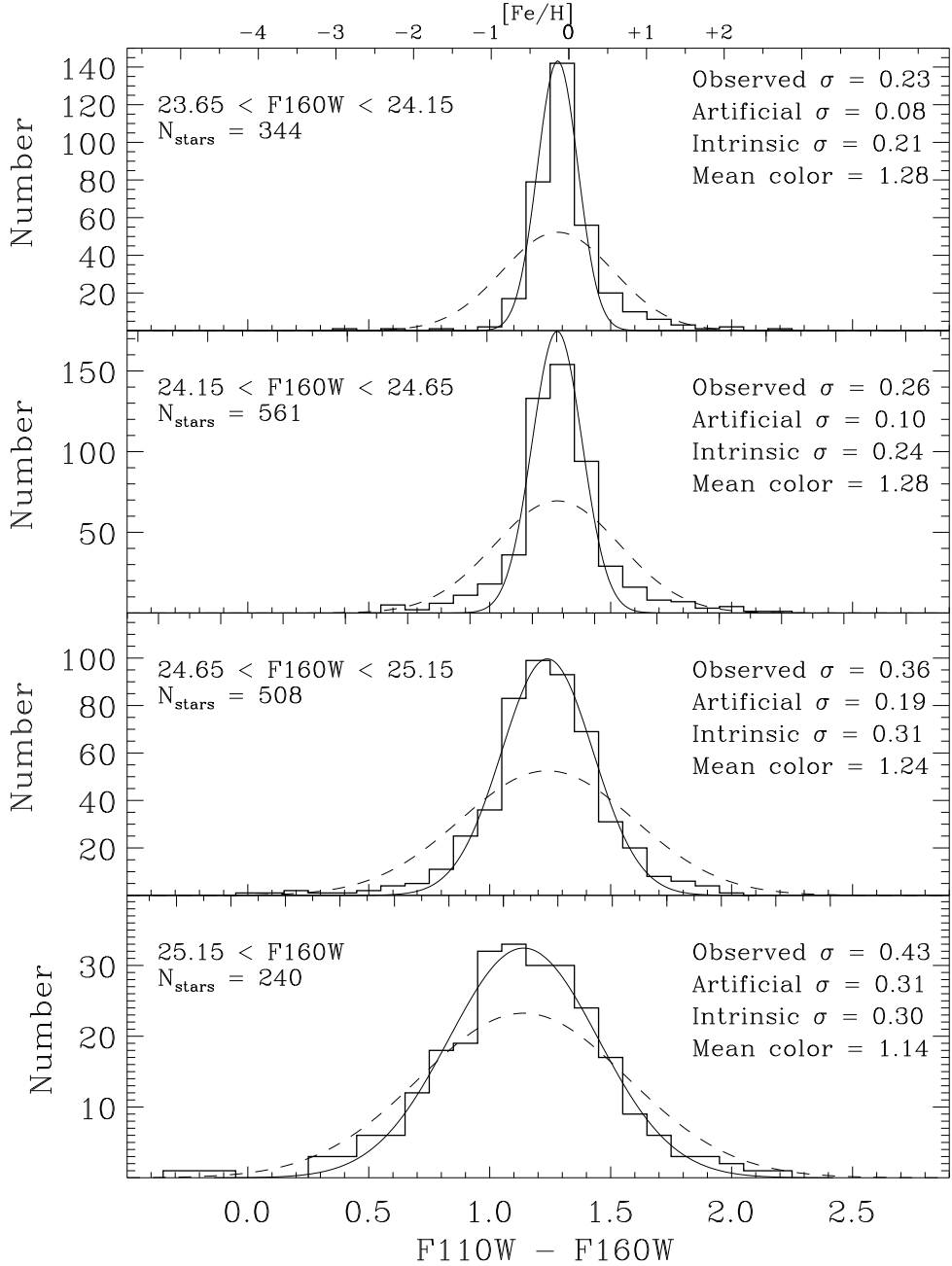}
\caption{\small A2: Histograms comparing the color spread of the
monochromatic artificial RGB (with $F110W-F160W=1.30$) to the real
stars at four magnitude intervals starting at the RGB tip.  The solid
line Gaussians are best fits to the plotted histograms; the dashed
line Gaussians have the same total area, but the widths of the real
star distributions in Figure~9.  The intrinsic color spreads are
estimated as the quadrature difference of the observed and
monochromatic artificial star color spreads.  }
\end{figure}

In Figure~A2, we compare the artificial data color spread to the
actual observed color spread, using the same magnitude intervals as in
Figure~12.  The quadrature difference in color spread between the real
and artificial images is our estimate of the intrinsic width of the
RGB.  The intrinsic color spread increases from 0.20 at the RGB tip to
0.30 at the limit of our data; this is consistent with the presence of
a very metal poor population entering at fainter magnitudes, as
predicted by the isochrones in a composite metallicity population
(c.f.\ Figure~11).  

The artificial RGB should have a mean color of exactly 1.3, but the
measured color moves 0.06 magnitudes bluer at $F160W=24.65$, where we
are 80\% complete, and 0.16 magnitudes bluer in the confusion-limited
regime.  The reddest faint stars go undetected in $F110W$, which causes
the shift to bluer colors and accounts for the difference between the
mean location of the observed RGB and the Girardi isochrones at faint
magnitudes (Figures~10, 11).

\clearpage
\begin{deluxetable}{cccccccccc}
\tablewidth{0pt}
\tablecaption{Artificial Star Photometry Tests}
\tablehead{
\multicolumn{2}{c}{Input}& \colhead{} & \multicolumn{7}{c}{Recovered
  (Input $F110W-F160W=1.1$)} \\ 
\cline{1-2} \cline{4-10}\\
\colhead{$F160W_{inp}$} &
\colhead{$F110W_{inp}$} &
\colhead{} &
\colhead{\%$_{rec}$} &
\colhead{$<color>$} &
\colhead{$\sigma_{color}$} &
\colhead{$<F160W>$} &
\colhead{$\sigma_{F160W}$} &
\colhead{$<F110W>$} &
\colhead{$\sigma_{F110W}$} 
}
\startdata
22.75 & 23.85 & &  99.7 & 1.102 & 0.061 & 22.750 & 0.053 & 23.848 & 0.055 \\ 
23.00 & 24.10 & &  99.7 & 1.104 & 0.077 & 22.998 & 0.068 & 24.096 & 0.069 \\ 
23.25 & 24.35 & &  99.8 & 1.102 & 0.091 & 23.243 & 0.084 & 24.340 & 0.086 \\ 
23.50 & 24.60 & &  99.4 & 1.106 & 0.115 & 23.493 & 0.101 & 24.589 & 0.102 \\ 
23.75 & 24.85 & &  99.0 & 1.107 & 0.143 & 23.735 & 0.128 & 24.836 & 0.132 \\ 
24.00 & 25.10 & &  97.4 & 1.111 & 0.172 & 23.977 & 0.156 & 25.073 & 0.162 \\ 
24.25 & 25.35 & &  94.8 & 1.113 & 0.217 & 24.214 & 0.194 & 25.310 & 0.198 \\ 
24.50 & 25.60 & &  88.3 & 1.121 & 0.248 & 24.427 & 0.237 & 25.519 & 0.239 \\ 
24.75 & 25.85 & &  77.3 & 1.156 & 0.280 & 24.609 & 0.276 & 25.709 & 0.293 \\ 
25.00 & 26.10 & &  63.0 & 1.198 & 0.323 & 24.735 & 0.328 & 25.861 & 0.366 \\ 
25.25 & 26.35 & &  51.8 & 1.230 & 0.326 & 24.762 & 0.429 & 25.921 & 0.421 \\ 
25.50 & 26.60 & &  45.4 & 1.231 & 0.321 & 24.695 & 0.547 & 25.895 & 0.517 \\ 
25.75 & 26.85 & &  43.2 & 1.246 & 0.312 & 24.634 & 0.580 & 25.837 & 0.560 \\ 
\hline
\multicolumn{2}{c}{Input}& \colhead{} & \multicolumn{7}{c}{Recovered
  (Input $F110W-F160W=1.5$)}\\
\cline{1-2} \cline{4-10}
22.75 & 24.25 & & 99.8 & 1.496 & 0.071 & 22.750 & 0.053 & 24.245 & 0.078  \\ 
23.00 & 24.50 & & 99.7 & 1.496 & 0.091 & 22.998 & 0.068 & 24.492 & 0.098  \\ 
23.25 & 24.75 & & 99.7 & 1.490 & 0.112 & 23.244 & 0.084 & 24.731 & 0.124  \\ 
23.50 & 25.00 & & 99.4 & 1.488 & 0.140 & 23.493 & 0.101 & 24.977 & 0.147  \\ 
23.75 & 25.25 & & 98.8 & 1.486 & 0.169 & 23.735 & 0.128 & 25.222 & 0.186  \\ 
24.00 & 25.50 & & 97.1 & 1.476 & 0.209 & 23.977 & 0.157 & 25.447 & 0.228  \\ 
24.25 & 25.75 & & 94.0 & 1.465 & 0.256 & 24.214 & 0.194 & 25.668 & 0.273  \\ 
24.50 & 26.00 & & 87.0 & 1.437 & 0.280 & 24.426 & 0.237 & 25.846 & 0.319  \\ 
24.75 & 26.25 & & 75.7 & 1.424 & 0.311 & 24.601 & 0.280 & 25.987 & 0.390  \\ 
25.00 & 26.50 & & 61.5 & 1.393 & 0.340 & 24.711 & 0.347 & 26.057 & 0.460  \\ 
25.25 & 26.75 & & 51.0 & 1.364 & 0.328 & 24.712 & 0.468 & 26.024 & 0.525  \\ 
25.50 & 27.00 & & 45.9 & 1.322 & 0.340 & 24.656 & 0.564 & 25.948 & 0.588  \\ 
25.75 & 27.25 & & 42.9 & 1.292 & 0.313 & 24.603 & 0.574 & 25.868 & 0.571  \\ 
%
\enddata
\end{deluxetable}

\clearpage

\section{Appendix B: NICMOS Color Transformations}

To compare NICMOS data with ground-based IR photometry for stars in
globular clusters and the Milky Way bulge requires a rather severe
transformation because $F110W$ is much wider and extends well to the
blue of standard $J$-band filters.  A number of transformations have
been published, but none were suitable for our purposes, either
because they did not reliably extend to cool giants (Origlia \&
Leitherer 2000; Marleau et al.\ 2000) or were based on outdated NICMOS
photometry zeropoints (Stephens et al.\ 2000).  

To derive an improved transformation, we used the Girardi isochrones
themselves, which are available in both ground-based Bessell-Brett
(1988) and NICMOS IR colors.  In keeping with our finding that the
stellar population of NGC3379 is relatively old and metal rich, and to
obtain the reddest possible points along a well-developed giant
branch, we base the transformation on the 14~Gyr isochrones of various
abundances.  Comparison of the isochrones in the two photometric
systems shows that there is little dependence of the transformation on
metallicity or age and also that the transformation relation clearly
differs from the linear approximation of Marleau et al.\ (2000) by up
to nearly 0.2 magnitudes for $J-H > 0.7$ (Figure~B1).  To derive our
improved transformation, we fit a high order polynomial to the
isochrone-isochrone comparison and then extended the transformation
linearly using the two reddest NICMOS photometric standards, OPH-S1
and CSKD-12 (Figure~B1).  Our reliance on a linear extrapolation for
colors redder than $F110W-F160W = 1.4$ is again probably an
oversimplification, but there are no additional data that can be
called into play.

\begin{figure}[t!]
\epsscale{0.8}
\plotone{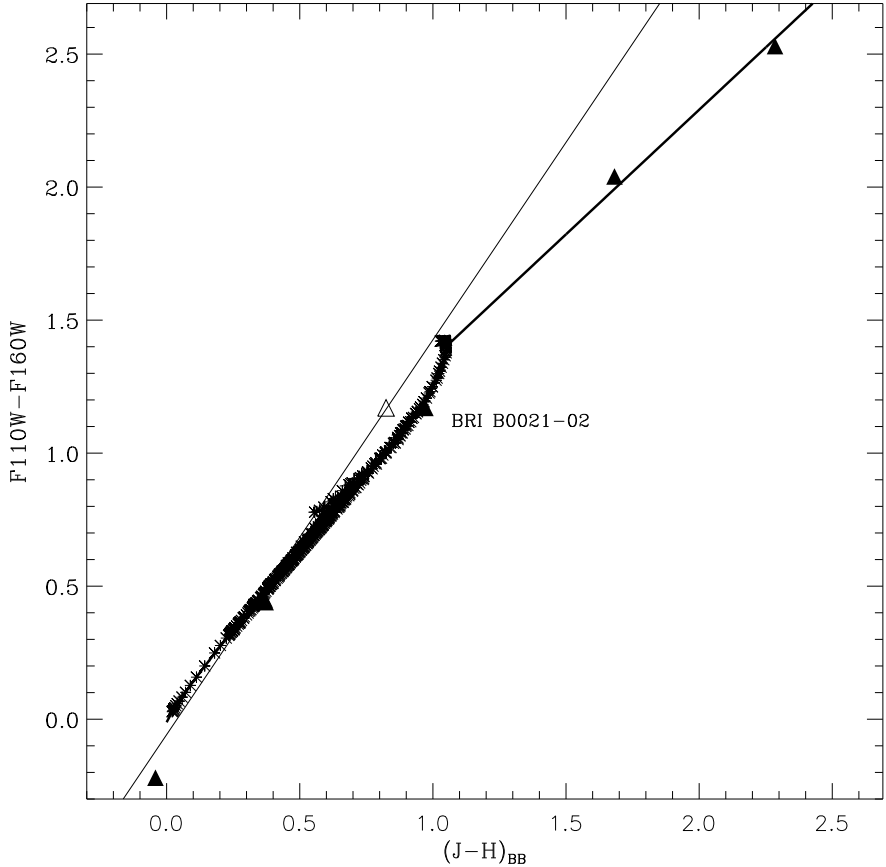}
\caption{\small B1: Ground-based Bessell-Brett $J-H$ to NICMOS $F110W-F160W$
transformation (heavy solid line) derived from the Girardi isochrones
for 14~Gyr and a range of abundances; from [Fe/H]$ = -2.27$ to +0.2,
there is little metallicity dependence of the photometric
transformation.  The individual isochrone points are plotted as
crosses; there are no points redder than what is shown.  The reddest
NICMOS photometric standards (filled triangles) are used to extend the
transformation linearly.  The M9.5V star BRI~B0021-02 is labeled; its
2MASS $J$ and $H$ magnitudes transformed to the Bessell-Brett system
(filled triangle) are in better agreement with the isochrones than the
photometry from the NICMOS STScI website (open triangle).  The
transformation adopted by Marleau et al.\ (2000) is shown as the light
solid line.  }
\end{figure}

Comparison with the four blue NICMOS standard star calibration data
($J-H < 1.$) shows that the isochrone-derived transformation yields a
good relation.  There has been some uncertainty concerning the correct
ground-based $J$-band photometry of the NICMOS standard BRI~B0021-02.
We have placed it in Figure~B1 using the $J-H$ color from the 2MASS
data base, transformed to the BB system using equations obtained at
the 2MASS website (http://www.ipac.caltech.edu/2mass).  The star's
location is consistent with the isochrones.  The ground-based
photometry listed at the NICMOS STScI website, however, places
BRI~B0021-02 about 0.15 magnitudes to the blue in the BB system (open
triangle in Figure~B1).  Marleau et al.\ (2000) use an even bluer
color for this star in deriving their linear transformation equation
(see their Figure~6 and Appendix), which perhaps led them to adopt a
color transformation considerably steeper than ours.

To test our transformation, we compared the IR photometry data of
Frogel et al.\ (1981) for 47~Tuc to the appropriate Girardi
isochrones.  The Frogel et al.\ data are on the CIT photometric
system, so it is necessary to apply the mild 10\% transformation from
CIT to the Bessell-Brett system given in Marleau et al.\ (2000).  The
left panel of Figure~B2 shows the resulting excellent fit of the
47~Tuc IR giant branch by the 14~Gyr, [Fe/H]$=-0.7$ Girardi isochrone,
adopting $m-M = 13.27$ for 47~Tuc; these numbers are consistent with
prevailing views of the stellar population of this cluster
(e.g. Hesser et al.\ 1987).  In the right panel, we have transformed
the cluster data points to the NICMOS system using our derived
transformation (Figure~B1) and over-plotted them on the same Girardi
isochrone for the NICMOS bandpasses.  Using our isochrone-derived
transformation, the 47~Tuc data, perhaps not surprisingly, are still
well-described by the same age and metallicity isochrone in the NICMOS
system.  This is not a trivial result, however; we have not applied
the transformation to the isochrones, only to the cluster data.  The
isochrones in the different photometric systems are produced by direct
convolution of theoretical spectral energy distributions with filter
transmission curves.  The transformation we have derived is in this
sense a wholly theoretical one, but it works well and is consistent
with the four NICMOS standards that overlap the isochrone color range.
Other published empirical ground-based--to--NICMOS photometric
transformations, however, result in large discrepancies of the 47~Tuc
data with the proper isochrone at these red colors.  The Marleau et
al.\ transformation, for instance, makes 47~Tuc appear to be solar
metallicity (Figure~B2), in conflict with all published results for
this cluster.

\begin{figure}[t!]
\epsscale{1.}
\plotone{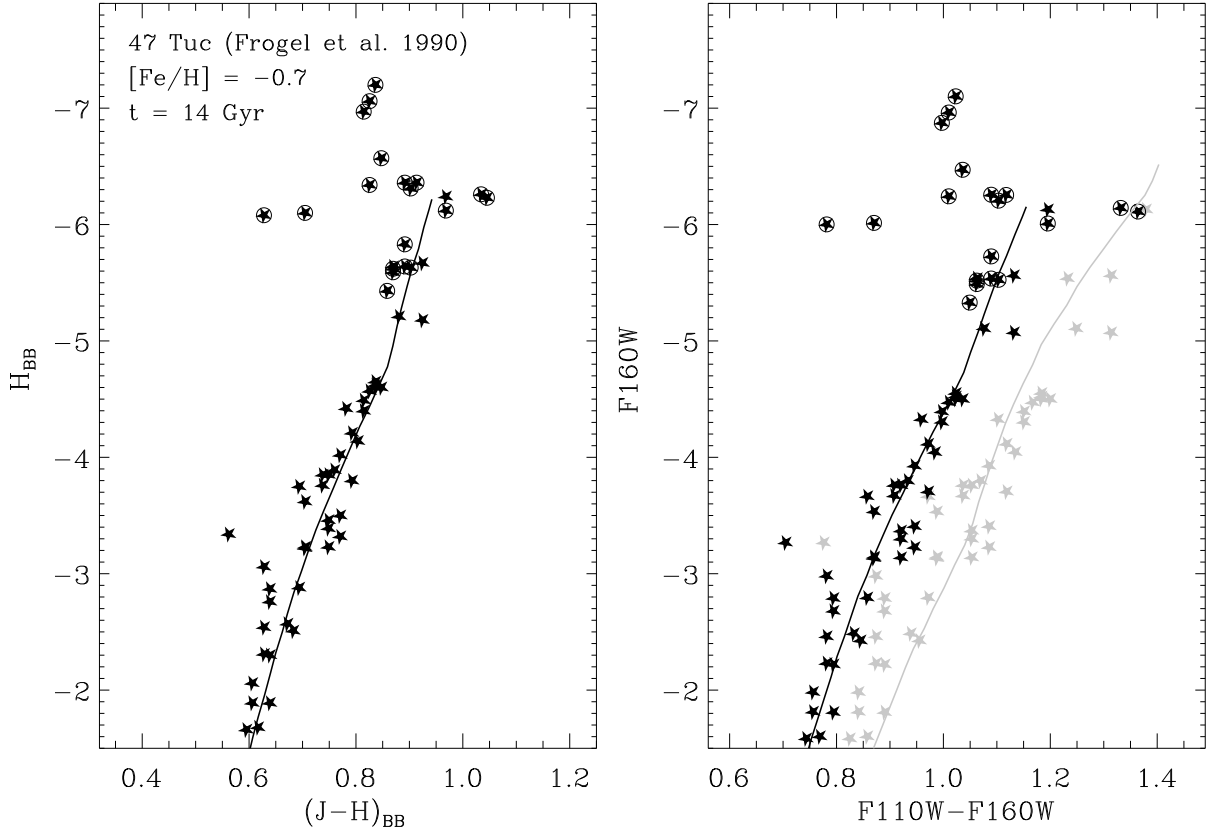}
\caption{B2: The left panel shows that the Girardi et al.\ 14~Gyr
isochrone for [Fe/H]$ = -0.7$ (solid line) is an excellent fit to the
infrared data of 47~Tuc from Frogel et al.\ (1990), after
transformation of the data to the Bessell-Brett (1988) system (stars;
known variables are circled).  In the right panel, the 47~Tuc points
have been transformed to the NICMOS system using the transformation
derived from the Girardi isochrones; the excellent agreement with the
isochrone is preserved.  Applying the transformation given by Marleau
et al.\ (2000), however, to transform the data to the NICMOS system
results in the RGB of 47~Tuc being too red (grey stars), making 47~Tuc
appear to match a 14~Gyr, solar abundance isochrone (grey line;
variables omitted for clarity).}
\end{figure}

\pagebreak

\pagebreak


\begin{references}
\reference{} Arimoto, N. 1996, in {\it From Stars to Galaxies}, ASP
Conference Series 98, Leitherer, C., Fritze-von Alvensleben, U., \&
Huchra, J., eds.
\reference{} Bekki, K., Couch, W. J., Drinkwater, M. J., \& Gregg, M. D. 2001,
             ApJL, 557, 39.
\reference{} Bertelli, G., Bressan, A., Chiosi, C., Fagotto, F., \&
  Nasi, E. 1994, A\&AS, 106, 275
\reference{} Blanco, B. M. 1984, AJ, 89, 1836
\reference{} Buta, R. J. \& McCall, M. L. 2003, AJ, 125, 1150
\reference{} Capaccioli, M., Held, E.V., Lorenz,
H. \& Vietri, M., 1990, AJ, 99, 1813
\reference{} Capaccioli, M., Vietri, M., Held, E. V., \& Lorenz,
H. 1991, ApJ, 371, 535
\reference{} Davidge, T. J. 2002, AJ, 124, 2012
\reference{} Davies, R. L., Burstein, D., Dressler, A., Faber, S. M., Lynden-Bell,
D., Terlevich, R. J. and Wegner, G. 1987, ApJS, 64, 581
\reference{} Davies, R. L., Sadler, E. M., \& Peletier, R. F. 1993,
MNRAS, 262, 650
\reference{} de Vaucouleurs, G. \& Capaccioli, M. 1979, ApJS, 40 699
\reference{} Elston, R. \& Silva, D. 1992, AJ, 104, 1360
\reference{} Faber, S. M.,  1973, ApJ, 179, 423
\reference{} Faber, S. M. et al.\ 1997, AJ, 114, 1771
\reference{} Freedman, W. L. 1989, AJ 98, 1285
\reference{} Freedman, W. L. 1992, AJ 104, 1349
\reference{} Freedman, W. L. et al.\ 2001, ApJ 553, 47
\reference{} Frogel, J. A., Persson, S. E., \& Cohen, J.G. 1981 ApJ,
             246, 842
\reference{} Frogel, J. A., Persson, S. E., \& Cohen, J.G. 1983 ApJS,
             53, 749
\reference{} Frogel, J. A. 1983 ApJ, 272, 167
\reference{} Frogel, J. A., Terndrup, D. M., Blanco, V. M., \& Whitford,
             A.E. 1990, ApJ, 353, 494
\reference{} Frogel, J. A. \& Whitford, A. 1987 ApJ, 320, 199
\reference{} Fruchter, A. S. \& Hook, R. N. 2002, PASP, 114, 144
\reference{} Gibson, B. K. et al.\ 2000, ApJ, 529, 723 
\reference{} Girardi, L., Bertelli, G., Bressan, A., Chiosi, C., 
             Groenewegen, M. A. T.,  Marigo, P., Salasnich, B., \&  
             Weiss, A. 2002, A\&A, 391, 195
\reference{} Graham, J. A. et al.\ 1997, ApJ, 477, 535
\reference{} Gregg, M. D. 1992, ApJ, 384, 43
\reference{} Gregg, M. D. 1995, ApJ, 443, 527
\reference{} Gregg, M. D. 1997, New Astronomy, 1, 363
\reference{} Grillmair, C. et al.\ 1996, AJ, 112, 1975
\reference{} Guarnieri, M. D., Renzini, A., \& Ortolani, S. 1997,
             ApJL, 477, 21
\reference{} Harris, G. L. H., Harris, W. E., \& Poole, G. B. 1999, AJ,
             117, 855
\reference{} Hesser, J. E., Harris, W. E., Vandenberg, D. A.,
             Allwright, J. W. B., Shott, P., \& Stetson, P. B. 1987
             PASP, 99, 739 
\reference{} Hjorth, J. \& Tanvir, N. R. 1997, ApJ, 482, 68
\reference{} Hughes, S. M. G., Wood, P. R. 1990, AJ, 99, 784
\reference{} Krist, J. 1993, in ASP Conf.~Ser.\ 52: Astronomical Data
             Analysis Software and Systems II, ed.\ R. J. Hanisch,
             R. J. V. Brissendon, \& J. Barnes, (San Francisco: ASP), 536
\reference{} Liu, M. C., Charlot, S. \& Graham, J. R. 2000, ApJ, 543, 644
\reference{} Leitherer, C., et al.\ 1996, in ASP Conf.~Ser.~98: From Stars to Galaxies, 
             ed.\ Leitherer, C., Fritze-von Alvensleben, U., \& Huchra, J.,
             (San Francisco: ASP)
\reference{} Luppino, G. A. \& Tonry, J. L. 1993, ApJ, 410, 81
\reference{} Maihara, T., et al.\ 2001, PASJ, 53, 25
\reference{} Maeder, A., Meynet, G. 1989, A\&A, 210, 155
\reference{} M\'endez, R. A.; Minniti, D. 2000, ApJ, 529, 911
\reference{} Minniti, D., Olszewski, E.W., Liebert, J. W., White, S. D. M.,
             Hill, J. M., \& Irwin, M. 1995, M.N.R.A.S, 277, 1293.
\reference{} Nieto, J.-L. \& Prugniel, P. 1987, A\&A, 186, 30
\reference{} Pastoriza, M. G., Winge, C., Ferrari, F., Macchetto, F. D.,
             \& Caon, N. 2000, ApJ, 529, 866
\reference{} Peletier, R. F. et al.\ 1999, MNRAS, 310, 863
\reference{} Rejkuba, M., Minniti, D., Silva, D. R., \& Bedding, T. R. 2001, A\&A, 
             379, 781
\reference{} Rejkuba, M., Minniti, D., Courbin, F., \& Silva, D. R. 2002, ApJ,
             564, 688
\reference{} Rejkuba, M., Minniti, D., \& Silva, D. R. 2003, A\&A, in
             press (astro-ph/0305432)
\reference{} Renzini, A. 1998, AJ, 115, 2459
\reference{} Rich, R. M.  1988, AJ, 95, 828
\reference{} Rose, J. A. 1985, AJ, 90, 1927
\reference{} Sadler, E. M., Rich, R. M., Terndrup, D. M. 1996, AJ,
             112, 171 
\reference{} Sadler, E. M. \& Gerhard, O. E. 1985, MNRAS, 214, 177
\reference{} Sakai, S., Madore, B. F., Freedman, W. L., Lauer, T. R.,
             Ajhar, E. A., \& Baum, W. A. 1997, ApJ, 478, 49
\reference{} Schlegel, D. J., Finkbeiner, D. P., \& Davis, M. 1998, ApJ, 500, 525
\reference{} Schneider, S. E. 1989, ApJ, 343, 94
\reference{} Soria, R., et al.\ 1996, ApJ 465, 79
\reference{} Schweizer, F. \& Seitzer P. 1992, AJ, 104, 1039
\reference{} Statler, T. S. 1994, AJ, 108, 111
\reference{} Statler, T. S. 2001, AJ, 121 244
\reference{} Statler, T. S. \& Smecker-Hane, T. 1999, ApJ, 117, 839
\reference{} Tanvir, N. R., Ferguson, H. C., Shanks, T., 
1999, MNRAS, 310, 175
\reference{} Terlevich, A. I. \& Forbes, D. A. 2002, MNRAS, 330, 547
\reference{} Thompson, R. I., Rieke, M.,  Schneider, G., Hines, D. C. \&
             Corbin, M. R.  1998, ApJL, 492, L95
\reference{} Tiede, G. P., Frogel, J. A., \& Terndrup, D. M. 1995, AJ,
             110, 2788
\reference{} Tonry, J. L., Ajhar, E. A., \& Luppino, G. A. 1990, AJ, 100, 1416
\reference{} Van den Bergh, S. 1989, PASP, 101, 1072
\reference{} Williams, R. E., et al.\ 1996, AJ, 112, 1335
\reference{} Worthey, G. \& Ottaviani, D. L.  1997, ApJS, 111, 377
\end{references}
\end{document}